\documentclass[fleqn,usenatbib,useAMS,usegraphicx,useAMS]{mnras}

\usepackage[T1]{fontenc}

\DeclareRobustCommand{\VAN}[3]{#2}
\let\VANthebibliography\thebibliography
\def\thebibliography{\DeclareRobustCommand{\VAN}[3]{##3}\VANthebibliography}

\usepackage{graphicx}	
\usepackage{amsmath}	
\usepackage{amssymb}	
\usepackage{breqn}
\usepackage{comment}
\usepackage{float}
\usepackage[normalem]{ulem} 

\def\euclid{{\it Euclid}}

\def\xx{\mathbf{x}}
\def\kk{\mathbf{k}}
\def\kka{\mathbf{k_1}}
\def\kkb{\mathbf{k_2}}
\def\kkc{\mathbf{k_3}}
\def\ka{\mathbf{k_a}}
\def\kb{\mathbf{k_b}}
\def\kc{\mathbf{k_c}}

\def\P{\mathcal{P}}

\def\ph{\mathbf{\hat{p}}}

\def\nn{\nonumber}

\def\fig#1{fig.~\ref{#1}}
\def\Fig#1{Figure~\ref{#1}}
\def\S#1{S_{#1}}

\def\bl{b_1}
\def\gn{\gamma_2}
\def\bto{\beta_1}
\defcitealias{Bharadwaj:2020wkc}{Paper~I}
\defcitealias{Mazumdar:2020bkm}{Paper~II}

\def\mpci{\,{\rm Mpc}^{-1}}

\title[]{Quantifying the redshift space distortion of the bispectrum III : 
Detection prospects of the multipole moments}

\author[A. Mazumdar, D. Sarkar, S. Bharadwaj]{ Arindam Mazumdar$^{1}$\thanks{\href{mailto:arindam.mazumdar@iitkgp.ac.in}{arindam.mazumdar@iitkgp.ac.in}}, 
Debanjan Sarkar$^{2}$\thanks{\href{mailto:debanjan@post.bgu.ac.il}{debanjan@post.bgu.ac.il}},
Somnath Bharadwaj$^{1,3}$\thanks{\href{mailto:somnath@phy.iitkgp.ac.in}{somnath@phy.iitkgp.ac.in}}   
\\
$^{1}$Centre for Theoretical Studies, Indian Institute of Technology Kharagpur, Kharagpur - 721302, India\\
$^{2}$Department of Physics, Ben-Gurion University of the Negev, Be’er Sheva - 84105, Israel\\
$^{3}$Department of Physics, Indian Institute of Technology Kharagpur, Kharagpur - 721302, India}

\date{}
\pubyear{2022}
\begin{document}
\label{firstpage}
\pagerange{\pageref{firstpage}--\pageref{lastpage}}
\maketitle

\begin{abstract}
The redshift space anisotropy of the bispectrum is generally
quantified using multipole moments. The
possibility of measuring these multipoles in any survey depends 
on the level of statistical fluctuations. We
present a formalism to compute the statistical fluctuations
in the measurement of bispectrum multipoles for galaxy surveys.
We consider specifications of a {\it Euclid} like galaxy survey
and present two quantities:
the signal-to-noise ratio (SNR) which quantifies the
detectability of a multipole, and the rank
correlation which quantifies the correlation in measurement
errors between any two multipoles. Based on SNR values, we find that
{\it Euclid} can 
potentially measure the bispectrum multipoles up to $\ell=4$
across various triangle shapes, formed by the three {\bf k}
vectors in Fourier space. In general, SNR is maximum for the linear triangles. SNR values
also depend on the scales and redshifts of observation. While, $\ell \leq 2$ multipoles
can be measured with ${\rm SNR}>5$ even at linear/quasi-linear ($k \lesssim 0.1 \mpci$) scales,
for $\ell>2$ multipoles, we require to go to small scales or need to increase  
bin sizes. For most multipole pairs, the errors are only weakly correlated 
across much of the triangle shapes barring a few in the vicinity of squeezed and stretched triangles.
This makes it possible to combine the measurements of different multipoles to increase 
the effective SNR. 

\end{abstract}



\section{Introduction}

Observations of the Cosmic Microwave Background (CMB)  
\citep{fergusson12-PNG,oppizzi18-PNG, Planck:2019kim, shiraishi19-PNG} and galaxy clustering 
\citep{feldman01-bispec, scoccimarro04-PNG, ligouri10-PNG, ballardini19png} indicate that the 
primordial  density fluctuations  are consistent with the the simplest models of inflation which   predict these  to be a Gaussian random field \citep{baumann09-inflation}. The power spectrum  is sufficient to quantify the statistics of a Gaussian random field for which  all the higher order statistics are predicted to be  zero.  However,  several inflationary scenarios also predict the primordial fluctuations to be non-Gaussian (primordial non-Gaussianity; \citealt{bartolo04-PNG-rev}).  Further,  the non-linear evolution  of  initially Gaussian  density 
fluctuations and also the non-linear biasing of the tracer fields (e.g. galaxies) both  introduce `induced non-Gaussianity' \citep{fry84-bispec, Bernardeau:2001qr}.
The bispectrum, which is the Fourier transform of the  three-point correlation function,  
is the lowest order statistics which is sensitive to the non-Gaussianity. 
Second order perturbation theory predicts   that measurements of the bispectrum in the weakly non-linear regime can be used to constrain the bias parameters \citep{matarrese97}, and this framework has been applied  in several  galaxy surveys to extract the galaxy bias parameters 
\citep{feldman01-bispec, Scoccimarro:2000sp, verde02-2DF, nishimichi07, Gil-Marin:2014sta}. 
Further, measurements of the bispectrum allow us to break the degeneracy between the  
matter density parameter $\Omega_m$ and the linear bias parameter $b_1$, 
something  which is not possible by using the power spectrum alone 
\citep{Scoccimarro:1999ed}.  A precise measurement of the Baryon Acoustic Oscillations (BAO)  in the bispectrum allows us to constrain the expansion rate of the universe
\citet{Pearson:2017wtw}. Upcoming galaxy surveys  like {\it Euclid} \citep{Euclid:2019clj}, promises to measure the bispectrum with high  precision,  and  thereby  constrain the
above parameters with unprecedented accuracy. 

Redshift space distortion (RSD) is an important effect in galaxy redshift surveys  
\citep{kaiser87,jackson72-FoG,hamilton-98-rsd-review}. The line-of-sight (LoS)  anisotrpy in the redshift space power spectrum  contains a wealth  of cosmological information. 
For example, this can  been  used to: $(i)$ measure the growth of structures $f$ on linear scales \citep{loveday96, peacock01, hawkins03, guzzo-pierleoni08}, 
$(ii)$ constrain the total density from   massive neutrinos
\citep{hu98-neutrino,Upadhye:2017hdl}, and   
$(iii)$ test dark energy and modified gravity theories 
\citep{linder08-RSD-GR, song09-RSD, de_la_torre16-gravity-test-from-RSD-and-lensing, johnson-blake-16-modified-gravity-using-galaxy-peculiar-velocities, mueller18-BOSS-RSD}. 

The  bispectrum also is affected by RSD, and the redshift space bispectrum  also contains a wealth  of cosmological information  \citep{Scoccimarro:1999ed,Yankelevich:2018uaz, Hahn:2020lou}, and it is   important to accurately model and quantify this. 
\citet{Hivon1995} and  \citet{Verde1998} have formulated the initial theoretical 
framework for calculating the bispectrum in redshift space. 
However,  they mainly focused on  measuring the large scale bias parameter and the cosmological parameters, and they have not quantified the RSD anisotropy in general.
Later, \citet{Scoccimarro:1999ed} have quantified the  anisotropy of the 
redshift space  bispectrum  utilizing  spherical harmonics.  
Their work, however, was restricted only to the monopole and one quadrupole component $(\ell=2,m=0)$.  The work of \citet{Hashimoto2017} also  was  limited to a single quadrupole component of  the redshift space bispectrum.  \citet{Nan} have calculated approximate analytical expressions for the  higher angular multipole moments up to $\ell=4$ based on the halo model, 
and their analysis is only limited   to a few  triangle configurations. 
\citet{Yankelevich:2018uaz} and \citet{Gualdi:2020ymf} have analysed the combined ability of the
redshift space  power spectrum and bispectrum to  constrain the cosmological parameters.
\citet{desjacques18-RSD-EFT} have  utilised effective field theory  to  model
the redshift space bispectrum on quasi non-linear scales.   The 
theoretical analysis of \citealt{Clarkson:2018dwn} and \cite{deWeerd:2019cae} suggests that 
relativistic effects will introduce a dipole anisotropy in the redshift space bispectrum 
on very large length-scales.  

There has been some work towards developing fast estimators which  quantify the anaistropy of the redshift space bispectrum.   \citet{Slepian:2016weg,Slepian2018} have introduced 
a technique to expand the redshift space three-point correlation function in terms 
of the products of two spherical harmonics. On the other hand,    \citet{Sugiyama2019a} have proposed  a tri-polar spherical harmonic decomposition to  quantify the anisotropy of the redshift space bispectrum,   and as a demonstration  they applied this  to the Baryon Oscillation Spectroscopic Survey (BOSS) Data Release 12.

Our recent work (\citealt{Bharadwaj:2020wkc}, hereafter  \citetalias{Bharadwaj:2020wkc})   presents a  formalism to quantify  the anisotropy of the redshift space bispectrum
$B^s(\ka,\kb,\kc)$ by decomposing it  into multipole moments $\bar{B}^m_{\ell}(k_1,\mu,t)$.  Here $k_1$, the length of the  largest side, and $(\mu,t)$ respectively quantify the size and shape of the  triangle $(\ka,\kb,\kc)$.   We have illustrated this formalism by  
quantifying  the anisotrpoy due to   linear RSD  of the bispectrum arising from primordial non-Gaussianity. We have found that only the first four even  $\ell$  multipoles are non-zero, 
for which we have presented  explicit analytical expressions. These  results are expected to be important  to  constrain $f_{\text{NL}}$ using the bispectrum measured from future  redshift surveys. In a subsequent work \citet{Mazumdar:2020bkm} (hereafter \citetalias{Mazumdar:2020bkm}),  the same formalism was used  to quantify the anisotropy of the induced redshift space bispectrum arising from Gaussin initial conditions. 
 We present  analytical expressions for all the  multipole moments  which are predicted to be non-zero ($\ell \le 8, m \le 6$ ) at  second order perturbation theory. 
 Considering triangles of all possible shapes, we  have analysed the shape dependence of all  the multipoles holding $k_1=0.2 \, {\rm Mpc}^{-1}, \bto=1, \bl=1$ and $\gn=0$ fixed.
Here  $\bto,\bl$ and $\gn,$   quantify the linear redshift distortion parameter, linear bias and  quadratic bias respectively.   For  most multipoles we find  that 
the maxima or minima, or both,  occur  very close to the squeezed limit. 
Further, the absolute values   $\mid \bar{B}^m_{\ell} \mid $ 
are found to decrease rapidly if either $\ell$ or $m$ are  increased. We have also provided rough estimates for  measuring  the various multipoles using upcoming galaxy redshift surveys.  The present paper presents a detailed analysis for the prospects of  measuring the various multipole moments using upcoming galaxy redshift surveys.

The ability to measure any particular  bispectrum multipole $\bar{B}^m_{\ell}$ primarily depends on its  amplitude, the galaxy number density  and  the extent  of the  survey volume. 
Until recently, measurements of the bispectrum were limited by the shot noise from the limited galaxy number density and the cosmic variance due to the limited  volume of the 
galaxy redshift surveys 
\citep{Scoccimarro:2000sp,verde02-2DF,2dFGRSTeam:2004jic,Jing:2003nb,Kulkarni:2007qu,Gaztanaga:2008sq,Marin:2010iv}. However,  \citet{Gil-Marin:2014sta,Gil-Marin:2016wya,Philcox:2021kcw} have recently measured the isotropic component (monopole) of the galaxy bispectrum
using the Baryon Oscillation Spectroscopic Survey at a relatively high level of precision. They  showed that it is possible to improve  the constraints on the cosmological  parameters by including  the bispectrum along with the power spectrum.
Upcoming galaxy surveys  will cover  unprecedented large volumes  with high galaxy
number density,  and it is anticipated that this  will enable us  to precisely measure the higher multipoles of the bispectrum.   For example, the {\it Euclid} space telescope \citep{Euclid:2019clj}
is expected to complete a wide survey 
that will measure $\sim \! 10^8$ galaxy redshifts over $15, 000$ square degrees on the sky
in the range $z \lesssim 2.5$. Precise measurements of the galaxy 
bispectrum using {\it Euclid} are expected to place  tight constraints on: 
$(i)$ primordial non-Gaussianity \citep{Fedeli:2010ud,Dai:2019tjh},
$(ii)$ neutrino masses \citep{Chudaykin:2019ock,Hahn:2020lou},
$(iii)$ modified gravity models \citep{Bose:2019wuz}, and 
$(iv)$ all the other currently constrained standard cosmological parameters \citep{Agarwal:2020lov}. The above estimates mostly rely on the real-space bispectrum or its redshift space  monopole. It is important and interesting  to also consider the   higher multipoles  of the bispectrum in order to utilize  the full reach of this mission \citep{Gualdi:2020ymf}.

In this paper we developed a formalism to calculate the  statistical fluctuations expected  in the bispectrum multipoles  measured from any   given galaxy redshift survey. We quantify these statistical  fluctuations through the error covariance of the different multipoles. Here  we have considered the specifications of the {\it Euclid} galaxy redshift survey for which we present the  signal-to-noise ratio (SNR) with which the different multipoles are expected to be measured. 
We also present results for the correlations expected between the different multipole moments.  
The paper is organised as follows. 
In Section~\ref{se:form}, we present the formalism
for calculating the error covariance between different bispectrum multipoles. 
In Section~\ref{se:results} we present our main results, which include the dependence of 
bispectrum multipoles on triangle shapes, sizes and redshifts, possibility of measuring the 
different multipoles, and error covariance between the pair of multipoles in terms of 
rank correlation.
Finally in Section~\ref{se:conclusion}, we summarize our findings and outline some future
directions.  
In this paper, we use the {\it Boltzmann} code \texttt{CLASS} (\citealt{Lesgourgues:2011re,Blas:2011rf}) 
to calculate the input matter power spectrum for our analysis and assumed
cosmological parameters from Planck 2015 ~\citep{planck-collaboration15} throughout 
the work.

\section{Formalism} \label{se:form}
\begin{figure}
    \centering
    \includegraphics[width=0.75\columnwidth]{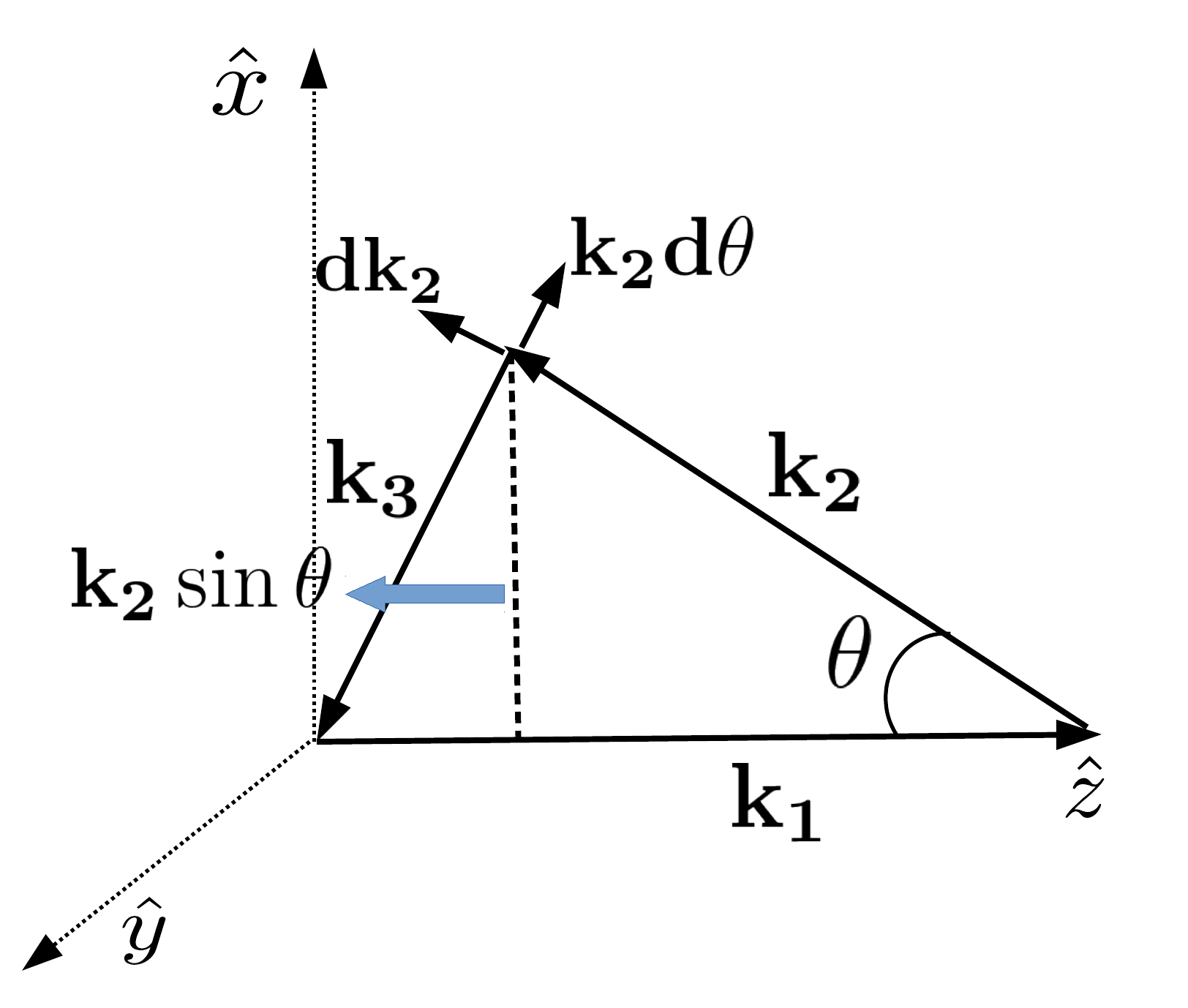}
    \caption{This shows a reference triangle ($\kka\kkb,\kkc$) in $x-z$ plane which is used to
    define various parameters used here. $\bf d k_2$, $d\theta$ represents the binning of the bi-spectrum triangle
    discussed in Section~\ref{subsec:binning}.}
    \label{fig:binning}
\end{figure}

We first consider   $\delta^s(\xx)$ which represents the density contrast  of a smooth  cosmological field   in redshift space. This  can be decomposed into  Fourier modes using 
\begin{equation}
  \delta^s(\xx)  = {1\over V} \sum_k e^{i \kk.\xx} \Delta^s(\kk)\, ,
\end{equation}
for which, in the absence of Poisson noise,  the power spectrum is defined as
\begin{equation}\label{eq:power}
    \langle \Delta^s(\kka)\Delta^s(\kkb) \rangle  = V \delta_{\kka+\kkb,0}P^s(k_1,\mu_1)\, ,
\end{equation}
where $\mu_1$ is the cosine of the angle between the Fourier mode  $\kka$ and the line of sight (LoS)  direction ${\bf \hat{n}}$. We have considered $ {\bf \hat{n}} =\hat{z}$ here.  
The bispectrum is  similarly  defined as
\begin{equation}\label{eq:bi}
 \langle \Delta^s(\kka)\Delta^s(\kkb)\Delta^s(\kkc) \rangle= V \delta_{\kka+\kkb+\kkc,0}B^s(\kka,\kkb,\kkc)  \,.
\end{equation}
Throughout this paper  we assume that the three  vectors $(\ka,\kb,\kc)$, which form a closed triangle,  are ordered such that $k_1 \ge k_2 \ge k_3$.  
In the absence of redshift space distortion the bispectrum  depends only on 
the shape and size of the triangle  formed by the three vectors $(\kka,\kkb,\kkc)$.  Here, following 
\citetalias{Bharadwaj:2020wkc},
we  parametrize  this using $(k_1,\mu,t)$   where $k_1$  the length of the  largest edge  quantifies the size of the triangle while  $\mu=\cos \theta$ (\fig{fig:binning}) and $t=k_2/k_1$ together quantity the shape of the triangle. Note that $\mu$ and $t$ are  both bounded  within the range $0.5 \le\mu,t\le 1$ and $2 \mu t \ge 1$.   In the presence of  redshift space distortion the bispectrum also depends on $\mu_a=\ka \cdot {\bf \hat{n}}/k_a$ where $a=1,2,3$. Therefore, the bispectrum now depends 
on how the triangle is oriented with respect to $ \hat{n} $. 

It is necessary to consider triangles of all possible orientations in order to quantify the anisotropy of the redshift space bispectrum.  Here we  start from a reference triangle  in the $x-z$ plane (\fig{fig:binning}) for which we have 
\begin{eqnarray}\label{eq:k_def}
\kka &=& k_1 \hat{z} \, , \nn \\
\kkb &=& k_1 t [- \mu  \hat{z} +  \sqrt{1-\mu ^2} \hat{x}]  \, ,
\end{eqnarray}
and $\kkc=-\kka-\kkb$. 
It is possible to obtain all possible orientations of the triangle by applying different rotations $\hat{\cal{R}}$ to  the reference triangle. We parameterize these rotations using $(\alpha,\beta,\gamma)$ the Euler angles which refer to successive rotations along  the $z$, $y$ and $z$  axes respectively  \citep{Sakurai:1167961}. 
We then have
\begin{eqnarray}
\mu_1&= &\ p_z  \,  \nn \\
\mu_2 &=&  -\mu \,  p_z  + \sqrt{1-\mu ^2} \, p_x \nn \\
\mu_3 &=& \frac{-[ (1-\mu  t) p_z   + t \sqrt{1-\mu ^2} p_x ] }
{\sqrt{1-2 \mu  t +t^2}}   
\label{eq:mu_rot}
\end{eqnarray}
where we have introduced an unit vector $\ph =  \hat{\cal{R}}^{-1} {\bf \hat{n}} $
which has components $p_z=\cos(\beta)$ and $p_x=-\sin(\beta) \, \cos(\gamma)$ respectively. 
 Note that these expressions are independent of $\alpha$ {\it i.e.} the axis of the  first rotation  ($z$)  coincides with ${\bf \hat{n}}$,  and it  does not affect the redshift space distortion.  
 In summary, the redshift space bispectrum $B^s(\kka,\kkb,\kkc)$ can be completely parameterized  using $B^s(k_1,\mu,t,\ph)$ where $(k_1,\mu,t)$ quantify the size and shape dependence of the triangle while $\ph$ quantifies its  orientation with respect to ${\bf \hat{n}}$.

The redshift space anisotropy of the bispectrum is quantified using multipole moments defined as 
\begin{equation}
\bar{B}^{m}_{\ell}(k_1,\mu,t) = \sqrt{\frac{(2 \ell +1)}{4 \pi}} 
\int [Y^m_{\ell}(\ph)]^{*}  B^s(k_1,\mu,t,\ph) \, d \Omega_{\ph}
\label{eq:bispec_int}
\end{equation}
where the integral over the solid angle $d \Omega_{\ph}$ accounts for all possible orientations of the triangle. As discussed in \citetalias{Bharadwaj:2020wkc}, only the even $\ell$ multipoles of the bispectrum are non-zero and these are all real valued. 

In reality, we encounter a finite number of Fourier modes distributed on a regular grid. 
Here we define  the bispectrum multipole estimator as a discrete sum over triangle configurations in Fourier space,  
\begin{equation}\label{eq:bispec_est}
\hat{B}_\ell^m (k_1,\mu, t) =   \sum_n \frac{w_l^m(\ph_n)}{2 V}  [\Delta^s({\kk}_n) \Delta^s(\bar{\kk}_n)
\Delta^s(\tilde{\kk}_n) +c.c.]
\end{equation}
where  $\kk_n+\bar{\kk}_n+\tilde{\kk}_n=0$ form a closed triangle,  each value of $n$ represents a different closed triangle, $c.c. \equiv$ complex conjugate,   and
\begin{equation}\label{eq:weight}
    w_\ell^m(n) = {\rm Re}\Big[\sqrt{2\ell +1\over 4\pi}{ Y_\ell^m(\hat{p}_n)
\over  \sum_{n_1} |Y_\ell^m(\ph_{n_1})|^2}\Big]\, .
\end{equation}
The sum over $n$ covers triangles of all possible orientations, with shape and size in a bin of extent $(\Delta k_1, \Delta \mu, \Delta t)$ around $(k_1,\mu,t)$. The exact binning scheme 
used for the present work is discussed later in this paper. 
\subsection{The Error Covariance}

We assume that the grid spacing in $\kk$ is adequately fine so that various triangle orientations are adequately sampled which implies that  $\langle \hat{B}_\ell^m (k_1,\mu, t) \rangle = \bar{B}^{m}_{\ell}(k_1,\mu,t)$ where $\langle ...\rangle$ denotes an ensemble average of different realisations of the random density field $\Delta(\kk)$. In the present work we are interested in the  statistical fluctuations (error) 
\begin{equation}
\Delta \hat{B}_\ell^m = \hat{B}_\ell^m(k_1,\mu, t)  - \bar{B}^{m}_{\ell}(k_1,\mu,t)
\label{eq:err}
\end{equation} 
which we quantify through the covariance 
\begin{equation}
    C_{\ell\ell'}^{mm'} = \langle \Delta \hat{B}_\ell^m \, \Delta \hat{B}_{\ell'}^{m'}  \rangle \,.  
    \label{eq:cov}
\end{equation}
We  calculate this using  
\begin{eqnarray}\label{cov_def_1}
&& C_{\ell\ell'}^{mm'} + \bar{B}^{m}_{\ell} \bar{B}^{m'}_{\ell'}
= (2 V)^{-2} \sum_{n_1, n_2} w_\ell^m(\ph_{n_1}) 
w_{\ell'}^{m'}(\ph_{n_2})   \times 
\nn \\
&& [ \langle  \Delta^s({\kk}_{n_1}) \Delta^s(\bar{\kk}_{n_1}) \Delta^s(\tilde{\kk}_{n_1})
  \Delta^s({\kk}_{n_2}) \Delta^s(\bar{\kk}_{n_2}) \Delta^s(\tilde{\kk}_{n_2}) \rangle +  \nn \\ && c.c. +  \nn \\ 
  &&\langle  \Delta^s({\kk}_{n_1}) \Delta^s(\bar{\kk}_{n_1}) \Delta^s(\tilde{\kk}_{n_1})
  \Delta^{s*}({\kk}_{n_2}) \Delta^{s*}(\bar{\kk}_{n_2}) \Delta^{s*}(\tilde{\kk}_{n_2}) \rangle  \nn \\ && + c.c. ] \,.
\end{eqnarray}
We have evaluated this using  
\begin{eqnarray}
&& \langle  \Delta^s({\kk}_{n_1}) \Delta^s(\bar{\kk}_{n_1}) \Delta^s(\tilde{\kk}_{n_1})
  \Delta^s({\kk}_{n_2}) \Delta^s(\bar{\kk}_{n_2}) \Delta^s(\tilde{\kk}_{n_2}) \rangle = \nn \\ 
 && [1+ 3 \delta_{n_1,n_2}] V^2   B^s(\kk_{n_1}, \bar{\kk}_{n_1},\tilde{\kk}_{n_1})   B^s(\kk_{n_2}, \bar{\kk}_{n_2},\tilde{\kk}_{n_2}) \nn \\
 && +  \delta_{n_1,-n_2} V^3 P^s(\kk_{n_1}) P^s(\bar{\kk}_{n_1})  P^s(\tilde{\kk}_{n_1}) 
 \label{eq:6delta}
\end{eqnarray}
which assumes that the signal is weakly non-Gaussian whereby contributiosn from $T^s$ the trispectrum and $P^s_6$ the sixth order polyspectrum can  be neglected. Here $\delta_{n_1,n_2}$ is a Kronecker delta which has value  one only when $n_1$ and $n_2$ refer to the same triangle and is zero otherwise. Further, $-n_2$ refers to the same triangle as $n_2$, but the two are exactly oppositely oriented {\it ie.} $\kk_{-n_1}=-\kk_{n_1}$, etc. Also note that $ \Delta^s({\kk}_{-n_2}) =\Delta^{s*}({\kk}_{n_2})$.  Using these in eq.~(\ref{eq:cov}) we obtain 
\begin{eqnarray}
&& C_{\ell\ell'}^{mm'} = \sum_{n} w_\ell^m(\ph_n) w_{\ell'}^{m'}(\ph_{n}) \times \nn \\ 
&& \{ 3 [B^s(\kk_{n}, \bar{\kk}_{n},\tilde{\kk}_{n})]^2 + 
 V P^s(\kk_{n}) P^s(\bar{\kk}_{n})  P^s(\tilde{\kk}_{n}) \}  
\label{eq:cov_def_2}
\end{eqnarray}
which provides an estimate of the error covariance in the absence of Poisson noise. 

\subsection{Poisson noise}
We now consider a discrete tracer ({\it e.g.} galaxies) whose density represents a Poisson sampling of the smooth filed which we have been discussing till now. We use  $\Delta_g(\kk)$ to represent the corresponding density contrast in Fourier space. 
The discrete sampling of the smooth field gives rise to a shot noise (or Poisson noise) 
whose contribution to the  power spectrum and the bi-spectrum are well studied  \citep{peebles_book,Scoccimarro:2000sp,smith_2009}. Using $\Delta_g(\kk)$ instead of $\Delta(\kk)$
 in eq.~(\ref{eq:power})  to estimate the power spectrum,  we have 
\begin{equation}
 V^{-1} \langle \Delta_g^s(\kka)\Delta_g^{s*}(\kka) \rangle  = P^s(k_1,\mu_1) + n_g^{-1}\, ,
\label{eq:pk}
\end{equation}
where, $P^s(k_a,\mu_a)$ is the power spectrum of the smooth  field $\Delta(\kk)$
and $n_g$ is the number density of galaxies. Similarly considering the bispectrum we have 
\begin{eqnarray}
&& V^{-1} \langle \Delta_g^s(\kka) \Delta_g^{s}(\kka) \Delta_g^{s}(\kkc) \rangle
  =  B^s(\kka,\kkb,\kkc) \nn +\\ 
&& n_g^{-1} [P^s(k_1,\mu_1)+P^s(k_2,\mu_2)+P^s(k_3,\mu_3)] + n_g^{-2} 
\label{eq:bk}
\end{eqnarray}
where it has been  assumed that $\kka+\kkb+\kkc=0$. It is now necessary to consider how eq.~(\ref{eq:cov_def_2}) for the bispectrum error covariance is modified if we account for  the discrete sampling. We first consider the ensemble average of the product of six $\Delta_g(\kk)$s which is required to calculate the covariance (eq.~\ref{eq:cov}).Here we use  the shortened notation $\Delta_{ag} \equiv \Delta_g(\ka)$,  $\Delta_a \equiv \Delta(\ka)$, $\Delta_{ab} \equiv \Delta(\ka+\kb)$, $\Delta_{abc} \equiv \Delta(\ka+\kb+\kc)$, etc.     We then 
have 
\begin{eqnarray}
&& \langle [ \Delta_{ag} \Delta_{bg}\Delta_{cg}\Delta_{dg}\Delta_{eg}\Delta_{fg}]
-[ \Delta_{a} \Delta_{b}\Delta_{c}\Delta_{d}\Delta_{e}\Delta_{f}] \rangle = \nn \\
&& V n_g^{-5} +  n_g^{-4} \langle [ \Delta_{a} \Delta_{bcdef}  +   \Delta_{ab} \Delta_{cdef} +  \Delta_{abc} \Delta_{cdef} ]  \rangle + \nn \\
&&  n_g^{-3} \langle [ \Delta_{a} \Delta_{b} \Delta_{cdef}  +   \Delta_{a}  \Delta_{bc} \Delta_{def} +  \Delta_{ab}  \Delta_{cd} \Delta_{ef} ]  \rangle + \nn \\
&&  n_g^{-2} \langle [ \Delta_{a} \Delta_{b} \Delta_{c} \Delta_{def}  +  
\Delta_{a} \Delta_{b} \Delta_{cd} \Delta_{ef}  ]  \rangle ] \nn \\
 && n_g^{-1} \langle [ \Delta_{a} \Delta_{b} \Delta_{c}  \Delta_{d}  \Delta_{ef} ] \rangle 
 + {\rm combinations} 
\label{eq:del6}  
\end{eqnarray}
where  all the terms in the {\it r.h.s.} arise due to the discrete sampling, and for each term it is necessary to also consider all the distinct combinations of the indices $a,b,c,...$.  Instead of embarking on a detailed calculation of the {\it r.h.s.}, we present it in a  schematic  form which can be used for an order of magnitude estimate. Using $[C_{\ell\ell'}^{mm'}]_g$ and $C_{\ell\ell'}^{mm'}$ to represent the error covariance respectively in the presence and absence of shot noise, we have 
\begin{eqnarray} 
&& [C_{\ell\ell'}^{mm'}]_g-C_{\ell\ell'}^{mm'} \sim  V^{-1} [ n_g^{-5} +  n_g^{-4} P^s +  n_g^{-3} B^s ] \nn \\
&& + n_g^{-2} [P^s]^2  + n_g^{-1} P^s B^s 
\label{eq:covg}
\end{eqnarray} 
where $C_{\ell\ell'}^{mm'} \sim V [P^s]^3 + 3  [B^S]^2 $ (eq.~\ref{eq:cov_def_2}).Here we consider estimates for the upcoming  \euclid{}  survey \citep{euclid2011}  which is expected to cover $V \sim 10^9 \, {\rm Mpc}^3$. Following \citet{Yankelevich:2018uaz} we adopt $n_g \sim  10^{-3}{\rm Mpc}^{-3}$ and a linear bias parameter $b\sim 1$ at $z=0.7$.  We have used 
\citet{Blas:2011rf} to estimate $P^s \sim 10^4 \, {\rm Mpc}^3$ and $B^s\sim [P^s]^2$ at $k \sim $ $0.2 {\rm Mpc}^{-1}$ where we expect second order perturbation to be reasonably valid. Using these,  we first consider  eq.~(\ref{eq:pk}) where we have  $P^s\sim 10^4 {\rm Mpc}^{3}$  and    $n_g^{-1} \sim 10^3 {\rm Mpc}^{3}$. Similarly in eq.~(\ref{eq:bk}) we have $B^s \sim 10^8  {\rm Mpc}^{6}$, whereas  $n_g^{-1} P^s \sim 10^7 {\rm Mpc}^{6}$ and $n_g^{-2}  \sim 10^6 {\rm Mpc}^{6}$. These estimates show  that  it is necessary to account for the shot noise in order to correctly estimate the power spectrum and the bispectrum. 
We now consider the error covariance for which we have $V [P^s]^3 \sim 10^{21} {\rm Mpc}^{12}$ and $[B^s]^2 \sim 10^{16} {\rm Mpc}^{12}$ whereby 
$C_{\ell\ell'}^{mm'} \sim  \sim 10^{21} {\rm Mpc}^{12}$ {\it i.e.} the error in the estimated bispectrum is dominated by the terms $ \sim V [P^s]^3$. Considering the shot noise, we see that we have the largest contribution from 
$n_g^{-1} P^s B^s \sim 10^{15} {\rm Mpc}^{12}$  and the magnitude decreases  with increasing power of $n_g^{-1}$ with the smallest contribution coming from 
$V^{-1} n_g^{-5} \sim 10^{6} {\rm Mpc}^{12}$. Even if we consider a lower galaxy number density 
$n_g \sim  10^{-4}{\rm Mpc}^{-3}$, we find that  the terms $V^{-1}[...] $ are all $\sim 10^{11} {\rm Mpc}^{12}$ and the two remaining terms are $\sim 10^{16} {\rm Mpc}^{12}$,  all  of which are several orders of magnitude smaller than the predicted value of $C_{\ell\ell'}^{mm'}$. We conclude that it is quite reasonable to ignore the   shot noise contribution, and we use  $C_{\ell\ell'}^{mm'}$ for the error estimates presented in the subsequent analysis. 

\subsection{Binning}\label{subsec:binning} 
We finally discuss the binning scheme which we have considered here. For the purpose of analytic predictions, it is convenient to assume that $V$ is very large whereby $d N_k$ the number of $\kk$ modes in the interval $d^3 k$ is given by $dN_k=(2 \pi)^{-3} V   \, d^3 k $.   The number of triangles $d N_{tr}$ is then given by 
\begin{equation}
dN_{tr} =  ( 2\pi)^{-6} V^2 \, d^3 k_1^3 \,  d^3 k_2^3\, ,    
\label{eq:dnk}
\end{equation}
and we replace the sum in eq.~(\ref{eq:cov_def_2})  using  $\sum_n \rightarrow \int dN_{tr}$. 
Starting from $\kka=k_1 \hat{x}$ (fig.~\ref{fig:binning}) we obtain all other possible vectors $\kka$ by changing the length $k_1$ or  rotations through the Euler angles $\alpha$ and $\beta$, and we have  
\begin{equation}
d^3 k_1 = k_1^2  \,   d k_1 \,  d\beta\, \sin\beta \,  d\alpha\,.
 \label{eq:dk1}
\end{equation}
Considering $\kkb$, a change in length could occur through either a change in $k_1$ or in $t$, whereas a change in orientation is associated with either a change in $\mu=\cos \theta$ or $\gamma$ the third Euler angle.  We then have 
\begin{equation}
    d^3 k_2= k_1^2  t^2 \, (t \, d k_1 + k_1 \, d t) \, d \mu \, d \gamma \,.
    \label{eq:dk2}
\end{equation}
 Here we have considered bins of extent $(\Delta k_1,\Delta \mu, \Delta t )$ centered around $(k_1,\mu,t)$.
 Using eq.~(\ref{eq:dk1}) and eq.~(\ref{eq:dk2}), we have 
\begin{equation}
d N_{tr}= ({8 \pi^2})^{-1}  N_{tr} \, d \alpha \, \sin \beta \, d \beta \, d \gamma
\label{eq:dnt}
\end{equation}
where  $N_{tr}$ the number of triangles in any particular bin is given by  
\begin{equation}
    N_{tr}=( 8 \pi^4)^{-1} \, (V k_1^3)^2 \, t^2 [ \Delta  \ln k_1 \, ( t \, \Delta  \ln k_1  + \Delta t)  \, \Delta \mu ]
    \label{eq:ntr}
\end{equation}

Using these we have 
\begin{equation}
\sum_{n_1} \mid Y^m_{\ell} (\ph_{n_1}) \mid^2=(4 \pi)^{-1} N_{tr}     
\end{equation}
which we use in the expression for $w_{\ell}^m$ (eq.~\ref{eq:weight}) to write 
eq.~(\ref{eq:cov_def_2}) as 
\begin{eqnarray}
&& C_{\ell\ell'}^{mm'}(k_1,\mu,t) =\frac{ \sqrt{(2 \ell+1)(2 \ell^'+1)}}{N_{tr} }
\int d \Omega_{\ph}  \times \nn \\
&& {\rm Re}[Y_{\ell}^m(\ph)] Re[Y_{\ell^'}^{m^'}(\ph)] \Big{\{} 3 [B^s(k_1,\mu,t,\ph)]^2  
\nn \\   
&& + V P^s(k_1,\mu_1) P^s(k_2,\mu_2) P^s(k_3,\mu_3) \,.
 \Big{\}} 
 \label{eq:cov_def_3}
\end{eqnarray}
Note that $k_2, k_3, \mu_1,\mu_2,\mu_3$ can all be calculated from  $(k_1,\mu,t,\ph)$
using eq.~(\ref{eq:k_def}) and eq.~(\ref{eq:mu_rot}). We have used eq.~(\ref{eq:cov_def_3}) to calculate the error estimates presented subsequently in this paper considering  bins of extent $(\Delta \ln k_1 = 0.1 ,\Delta \mu = 0.05, \Delta t = 0.05)$.

\subsection{Second Order Induced Bispectrum}
The induced bispectrum for any tracer in redshift space from second-order perturbation theory 
(2LPT) can be written as 
\begin{dmath}\label{eq:bispec}
B_{\rm 2PT}^s(\kka,\kkb,\kkc) = 2  \bl^{-1}(1+\bto\mu_1^2)(1+\bto\mu_2^2)\Big\{ F_2(\kka,\kkb)+{\gn\over 2}
+ \mu_3^2\bto G_2(\kka,\kkb)-\bl \bto \mu_3 k_3 \Big[ {\mu_1\over k_1}(1+\bto\mu_2^2)
+ {\mu_2\over k_2}(1+\bto\mu_1^2)\Big]\Big\} P(k_1) P(k_2) + 
{\rm cyc ...}\, ,
\label{eq:zb1}
\end{dmath}
where $\bto$  is the linear redshift distortion parameter, $\gamma_2 = b_2/b_1$ where $b_1$ is the linear bias
and $b_2$ is the quadratic bias, and 
\begin{equation}
G_2(k_1,k_2)={3\over 7}+{\kka\cdot \kkb\over 2}\left({1\over k_1^2}+{1\over k_2^2}\right)+{4\over7}{(\kka \cdot \kkb)^2\over k_1^2 k_2^2},
\label{eq:G2}
\end{equation}
refers to the second order kernel for the divergence of the peculiar velocity. Here we find it convenient to use
the following notation 
\begin{equation}
F_2(\kka,\kkb) = G_2(\kka,\kkb) + \Delta G(\kka,\kkb)  
\label{eq:F2}
\end{equation}
where 
\begin{equation}
\Delta G(\kka,\kkb)= {2\over 7}\left[1-{(\kka \cdot \kkb)^2\over k_1^2k_2^2}\right]\,.
\label{eq:dG2}
\end{equation}
Note that, the notations used here are taken explicitly from \citetalias{Bharadwaj:2020wkc} 
and \citetalias{Mazumdar:2020bkm}.

In \citetalias{Mazumdar:2020bkm},  we have quantified the anisotropy of the induced redshift-space 2LPT
bispectrum (eq.~\ref{eq:zb1}) in terms of the multipole moments $\bar{B}^{m}_{\ell}(k_1,\mu,t)$ 
defined in eq.~(\ref{eq:bispec_int}). There we have also presented  the formulas needed to
calculate all the multipole moments that are predicted to be non-zero at second-order
perturbation theory. Further, we have analysed the $\mu-t$ dependence of $\bar{B}^{m}_{\ell}(k_1,\mu,t)$ 
at fixed $k_1=0.2\,\mpci$ and for the parameter values $\beta_1=1$, $b_1=1$, $\gamma_2=0$. 
We find that the even multipole moments $\bar{B}^{0}_{0}$, $\bar{B}^{0}_{2}$ and $\bar{B}^{0}_{4}$
show very similar behaviour where their values are positive in the whole $\mu-t$ plane, the 
smallest values occur near the equilateral triangles and the largest values are found for the 
linear triangles. Two other even multipole moments $\bar{B}^{2}_{2}$ and $\bar{B}^{4}_{4}$
show positive values in the whole $\mu-t$ plane, however, their shape dependence is different from the multipoles discussed above. 
For the rest of the multipole moments, we find that they exhibit negative values
at some parts of or through the entire $\mu-t$ plane. Their contour patterns are also 
different from each other. We broadly see that, although the higher multipoles ($\ell>2$) 
show rich variety of shape dependence, their amplitude $\mid \bar{B}^{m}_{\ell} \mid$
fall off sharply as $\ell$ and $m$ are increased. Note that, the analysis in
\citetalias{Mazumdar:2020bkm} ignores the Finger-of-God (FoG) effect.  At large scales linear perturbation theory  and its second order extension  are expected to provide   a reasonably  accurate description of   the  clustering of matter in real space. This however does not hold in redshift space where it is found that  large peculiar velocities arising from highly non-linear small scale structures cause the large scale clustering pattern to appear elongated along the LoS. This is  known as the Finger of God (FoG) effect, and it is important to include this in  any realistic analysis of the redshift space power spectrum and bispectrum. 

\subsection{Finger of God Effect}
In case of the power spectrum, the FoG effect is generally incorporated  as an ad-hoc damping profile which multiplies the linear redshift space power spectrum.
A number of profiles, namely
Lorentzian, Gaussian, Lorentzian-squared {\it etc.}, have been considered in the literature. Although the 
Lorentzian profile works well for the simulated data 
\citep{Davis:1982gc, Hamilton:1997zq, Hatton:1999rs, Seljak:2000jg, White:2000te, Sarkar:2018gcb, Sarkar:2019nak}, 
the Gaussian profile is expected to occur naturally
\citep{bharadwaj01, scoccimarro04, hikage13, hikage13b, okumura15, hikage16}. 
In this work, we consider a Gaussian profile for the FoG damping and model the FoG  
redshift-space power spectrum  as \citep{peacock_book} 
\begin{equation}\label{eq:pk_FoG}
    P^s_{\rm FoG}(k_1,\mu_1) = 
    \exp{\left[(-k_1^2\mu_1^2){\sigma_p^2\over 2}\right]}  \times P^s_{\rm 1PT}(k_1,\mu_1)
\end{equation}
where $P^s_{\rm 1PT}(k_1,\mu_1) =(1+\beta_1 \mu_1^2)^2 \, P(k_1)$ is the linear redshift space power spectrum,  and  $\sigma_p$ parametrizes the pairwise velocity dispersion 
 is in units of comoving ${\rm Mpc}$. We can equivalently use
$[\sigma_p \, a \, H(a)]$  in units of $ {\rm km \, s^{-1}}$.
Note that on very large scales $k_1 \sigma_P \ll 1$, the damping factor $ \exp{\left[-k_1^2\mu_1^2{\sigma_p^2/2 }\right]} \approx 1$, and the  
 Kaiser effect $(1+\beta_1 \mu_1^2)^2$
 suffices  to describe the RSD effect.

We similarly  incorporate the FoG effect for the bispectrum as
\begin{eqnarray}\label{eq:B_FoG}
B^s_{\rm FoG}(\kka,\kkb,\kkc)&=&  \exp \left[-(k_1^2\mu_1^2+k_2^2\mu_2^2+k_3^2\mu_3^2){\sigma_p^2\over 2}\right] \nn \\ &&
\times 
B_{\rm 2PT}^s(\kka,\kkb,\kkc)
\label{eq:Bsp_rsd}
\end{eqnarray}
where we use eq.~(\ref{eq:zb1}) to calculate $B_{\rm 2PT}$. Due to the FoG damping term, we do not have closed form analytic expressions  for the 
various  multipole moments  of $B^s_{\rm FoG}$. Here we have  
numerically integrated eq.~(\ref{eq:bispec_int})  to compute the various multipole moments for which the  results are presented in the following section. 

\section{Results}\label{se:results}
In this section we present the shape and size dependence of the redshift space bispectrum,  and also the corresponding statistical errors for an \euclid{} like survey.  As discussed in Section \ref{se:form}, we have used  the largest side  $k_1$  to quantify the size of the triangle, and we have used $\mu=\cos \theta$ and $t=k_2/k_1$ (\fig{fig:binning}) to quantify the shape.  Following \citet{fry84-bispec},  we have defined  the  dimensionless bispectrum multipoles  ~ \citepalias{ Mazumdar:2020bkm})   
\begin{equation}
    Q_\ell^m(k_1,\mu, t) = b_1 {\bar{B}^{m}_{\ell}(k_1,\mu,t)\over 3 P(k_1)^2}\, ,
\end{equation}
which we have used to exhibit the results presented here.

Galaxy surveys like Euclid expect to cover a broad  redshifts range  from $0.001$ to $2.5$~\citep{Euclid:2021qvm,Pozzetti:2016cch}, which can be divided into various redshift bins for the subsequent analysis.  For the present work,  we  use  the $z$ bins as used in \citet{Yankelevich:2018uaz}  and we also adopt their predicted values for  the bias parameters $b_1$ and $\gamma_2$, the survey volume $V$ and  the pairwise velocity dispersion $\sigma_p$. 
We  have used the Boltzmann code CLASS~\citep{Lesgourgues:2011re,Blas:2011rf} to compute the real space matter power spectrum and the growth rate $f$ for each $z$ bin.

We have considered the $z=0.7$ redshift bin as the fiducial value for which most of our results are shown.   Following \citet{Yankelevich:2018uaz}, we have adopted the values
$b_1=1.18$, $\gamma_2=-0.9$,  $V=8.97 \,{\rm Gpc}^3$ and $\sigma_p = 7.09$Mpc for this particular $z$ bin.  \Fig{fig:qlm}  shows $Q_\ell^m$ as functions of $\mu$ and $t$  at $k_1 = 0.2 {\rm Mpc^{-1}}$  for the fiducial redshift $z=0.7$. Here  $k_1 = 0.2 {\rm Mpc^{-1}}$ is  a sufficiently  large length-scale where we may expect a combination of 2PT and  FoG  to provide a reasonably good description of the redshift space bispectrum. The cosmic variance is expected to increase if we consider small values of $k_1$, whereas non-linear effects increase if we consider larger $k_1$. Guided by this, we have mainly shown the results for  $k_1 = 0.2 {\rm Mpc^{-1}}$.  
We now briefly discuss the shapes of the triangle corresponding to different values of $\mu$ and $t$, the reader is referred to \citetalias{Bharadwaj:2020wkc} for further details. Considering any panel of \fig{fig:qlm}, the right boundary $\mu=1$ corresponds to linear triangles where the three sides are co-linear. Here, the bottom right  corner  $(\mu,t)=(1,0.5)$ corresponds to stretched triangles where $k_2=k_3=k_1/2$, and the top  right corner $(1,1)$ corresponds to squeezed triangles where $k_1=k_2,k_3 \rightarrow 0$. The top boundary $t=1$ corresponds to  L-isosceles triangles where the two larger sides have equal length $(k_1=k_2)$, and the bottom boundary $2 \mu t =1$ corresponds to S-isosceles triangles where the two  smaller sides have equal length $(k_2=k_3)$. The top left  corner $(0.5,1)$ corresponds to equilateral triangles. The diagonal $\mu=t$ corresponds to right-angled triangles, and the regions  $t>\mu$ and $t < \mu$ correspond to acute  and obtuse triangles respectively.

\subsection{Predicted Multipole Moments}
\begin{figure*}
    \centering
   \includegraphics[width=0.98\textwidth]{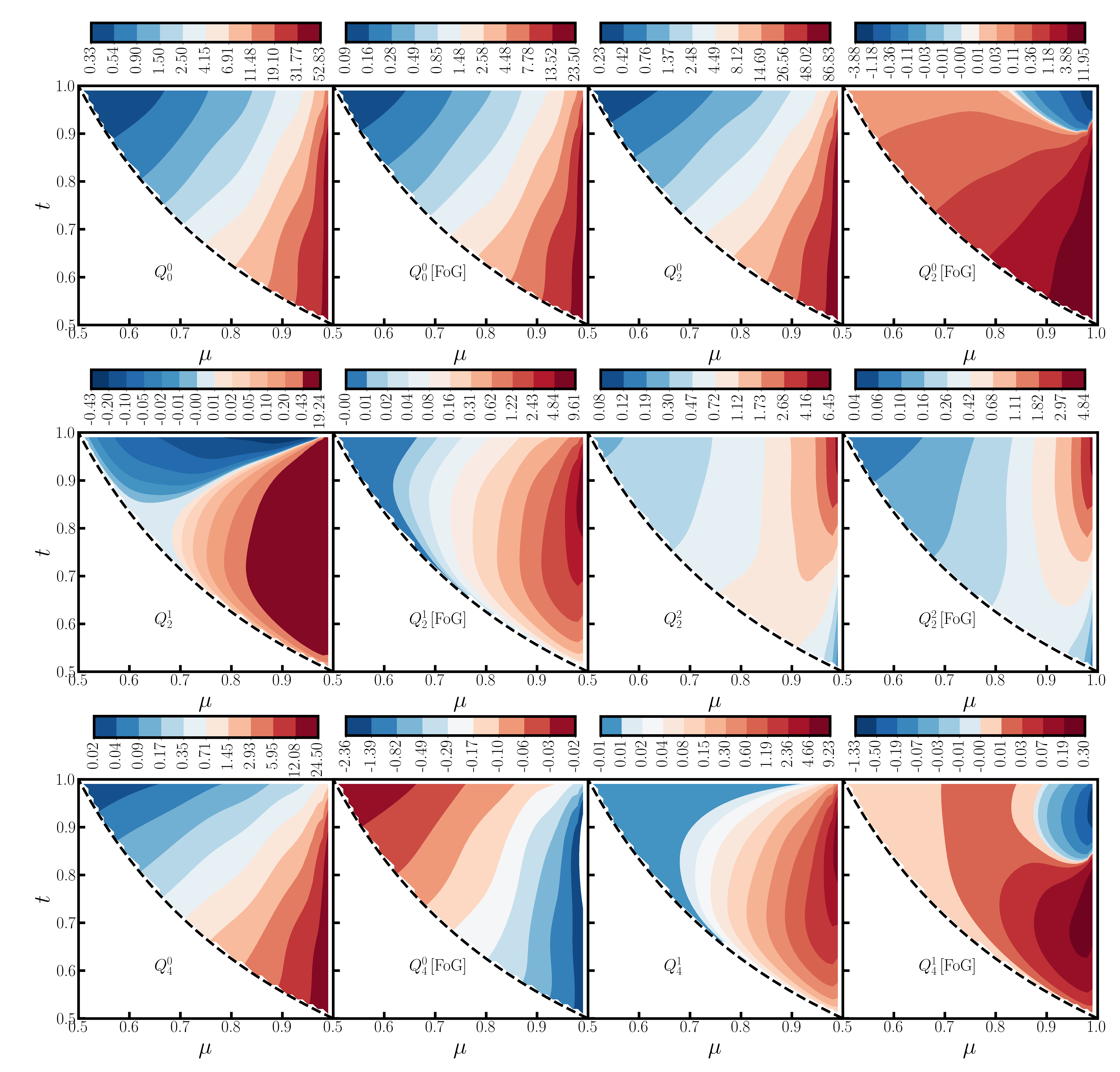}
    \caption{This shows $Q_\ell^m$ as functions of $\mu$ and $t$  at $k_1 = 0.2 {\rm Mpc^{-1}}$  for the fiducial redshift $z=0.7$. $Q_\ell^m$[FoG] means that the effect of FoG has been considered in bispectrum according to eq.~(\ref{eq:B_FoG}). In other plots, where FoG is not mentioned in the parenthesis, the bispectrum is considered only up to the 2PT level 
    (eq.~\ref{eq:bispec}). From the plots it is evident that FoG suppress the bispectrum significantly at $k_1=0.2{\rm Mpc}^{-1}$.}
    \label{fig:qlm}
\end{figure*}

In \fig{fig:qlm}, we have shown   $Q_\ell^m$  for $\ell=0,2$ and $4$.  As mentioned earlier,  
only the even $\ell$ multipoles  are expected to be non-zero. Further, in 2PT we  
expect non-zero multipoles up to $\ell=8$ \citepalias{Mazumdar:2020bkm}, however the 
value of  $Q_\ell^m$ falls with increasing $\ell$ and $m$, and we have shown the main results 
only up to  $\ell=4,m=3$.  As shown later in this  section,  we do not expect to have a 
statistically significant measurement of  the  higher  multipoles for the survey parameters 
considered here.  Note that 
$Q_\ell^m$   and $Q_\ell^m$[FoG]  respectively denote the results without and with the FoG. In all cases 
we find that the magnitude of $Q_\ell^m$[FoG]s is  highly suppressed compared to $Q_\ell^m$ due to the FoG effect.  Considering $Q_0^0$, we find that the  the minima  $(0.33)$ occurs for  the equilateral 
triangles and the the maxima  (52.83) occurs close to the squeezed limit ($\mu\rightarrow 1, t = 0.865$). 
The  minimum and maximum values fall to  $0.09$ and $23.50$  for $Q_0^0$[FoG]. We notice that the drop in value due to the FoG is more pronounced for equilateral triangles where the minima occurs. However, the overall patterns are visually very similar for $Q_\ell^m$   and $Q_\ell^m$[FoG].

The contour pattern of $Q_2^0$ shows  features similar to those of  $Q_0^0$. Its peak value ($86.83$) is larger than the peak 
value of $Q_0^0$ and it occurs at $\mu =0.995, t =0.795$. We observe that the FoG effect causes a larger  suppression
for the  quadrupole than in monopole, and  the maximum value of $Q_2^0$[FoG] drops to  11.95 which  occurs at a slightly different location (at $\mu = 0.995, t=	0.735$). It is interesting to note that in the absence of  the FoG damping $Q_2^0$ is larger than than  all the other multipoles including the monopole. However as soon as we apply the  FoG effect, the peak value of $Q_2^0$[FoG] drops and and  $Q_0^0$[FoG] overtakes  it  to become the multipole with the  largest value. The entire contour pattern of $Q_2^0$[FoG]  is quite different from that of  $Q_2^0$.  Unlike $Q_2^0$ which is positive everywhere, we see that $Q_2^0$[FoG] is negative over a region near the squeezed limit. The minimum  values of $Q_2^0$ (0.23) and $Q_2^0$[FoG] (-3.88) respectively occurs at the equilateral and squeezed limits, however the magnitude of  $Q_2^0$[FoG] is still  minimum for equilateral triangles.

Considering $Q_2^1$, we find that this has  negative values for most of the acute triangles and positive values for obtuse triangles, with a zero crossing somewhat above the line $\mu=t$ which corresponds to right-angled triangles. The minima (-0.43) and maxima  (19.24) 
respectively occur  at ($\mu = 0.965, t\rightarrow 1$) and ($\mu \rightarrow 1, t = 0.915$) which are both very close to the squeezed limit. The FoG effect substantially changes the contour pattern,    and we see that  $Q_2^1$[FoG] is positive valued  throughout. The maxima (9.61)  now occurs near the squeezed limit while the minima ($\approx 0$) occurs for equilateral triangles.

In case of $Q_2^2$, we find that the minimum value (0.08) is near the stretched limit and 
the maximum value (6.45) occurs close to the squeezed limit ($\mu\rightarrow 1, t = 0.975$). 
The maximum (4.84) and minimum (0.04) values in case of $Q_2^2$[FoG] do not change position, 
only the amplitudes drop. 
Overall, $Q_2^2$ and $Q_2^2$[FoG] show similar contour patterns.

Considering $Q_4^0$, we see that it shows similar shape dependence as $Q_0^0$ and $Q_2^0$, 
only its maximum (21.5) and minimum (0.02) values are smaller. Comparing 
$Q_4^0$[FoG] with $Q_4^0$,  we see that the patterns are  visually similar, however the values of  $Q_4^0$[FoG] are all negative and they are roughly an order of magnitude smaller. Considering $Q_4^1$, we see that  its  minima (0.01) occurs  for  equilateral triangles and the maxima (9.23) is 
at ($\mu\rightarrow 1, t =0.905$) which is somewhat below the squeezed limit.  Comparing  $Q_4^1$[FoG] with $Q_4^1$, we see that the patterns are quite different,  and the values of  $Q_4^1$[FoG] are roughly an order of magnitude smaller. We see that  $Q_4^1$[FoG] is negative near the squeezed limit  where  $\mid Q_4^1 \mid$[FoG] has a maxima (1.33), whereas the minima (0.30) occurs for equilateral triangles.   In summary we note that   the FoG effect has a significant impact on the redshift space bispectrum at the length-scales and redshifts of our interest, and we have  included  this in our subsequent analysis.

It is interesting to compare 
the  $Q_\ell^m$ values (without FoG) shown in \fig{fig:qlm} with those 
shown in \citetalias{Mazumdar:2020bkm}. We note that the bias parameters and redshift $z=0.7$ used here are different from those used in \citetalias{Mazumdar:2020bkm}. 
 We find that the contour patterns of  $Q_0^0, Q_2^0$ and $Q_4^0$  look very similar  in both \fig{fig:qlm} and \citetalias{Mazumdar:2020bkm},   however   their values are $\geqslant 5$ times larger in \citetalias{Mazumdar:2020bkm} which considers $z=0$.  On the other hand, the odd multipoles are found to be very different.  For example, in \citetalias{Mazumdar:2020bkm}  the minima and maxima of  $Q_2^1$ lie along the $\mu=1$ and $t=1$ lines respectively   whereas  it is exactly the opposite in \fig{fig:qlm}.   This is also true for $Q_4^1$ where the contour patterns show completely opposite trends in \citetalias{Mazumdar:2020bkm} and \fig{fig:qlm}. These differences are  mainly due to the choice of the non-linear bias parameter
$\gamma_2$ which is set to  zero in \citetalias{Mazumdar:2020bkm} whereas we have used   $\gamma_2=-0.9$  for  \fig{fig:qlm}.

The entire discussion has been restricted to $k_1= 0.2 {\rm Mpc}^{-1}$ till now. It is worth noting that the  
impact  of the FoG effect falls exponentially as we move to lower $k$ (large scales), and  it  causes $\lesssim 5\%$change in the bispectrum at $k_1= 0.05{\rm Mpc^{-1}}$ for $z=0.7$. Second order perturbation theory (eq.~\ref{eq:zb1}) alone is adequate  to model the redshift space bispctrum at  $k_1< 0.05{\rm Mpc^{-1}}$,   however  we do not consider these small $k_1$ values in our work as most of the bispectrum multipoles become undetectable due to cosmic variance (as shown later). 

\subsection{SNR predictions}
\begin{figure}
    \centering
    \includegraphics[width=0.97\columnwidth]{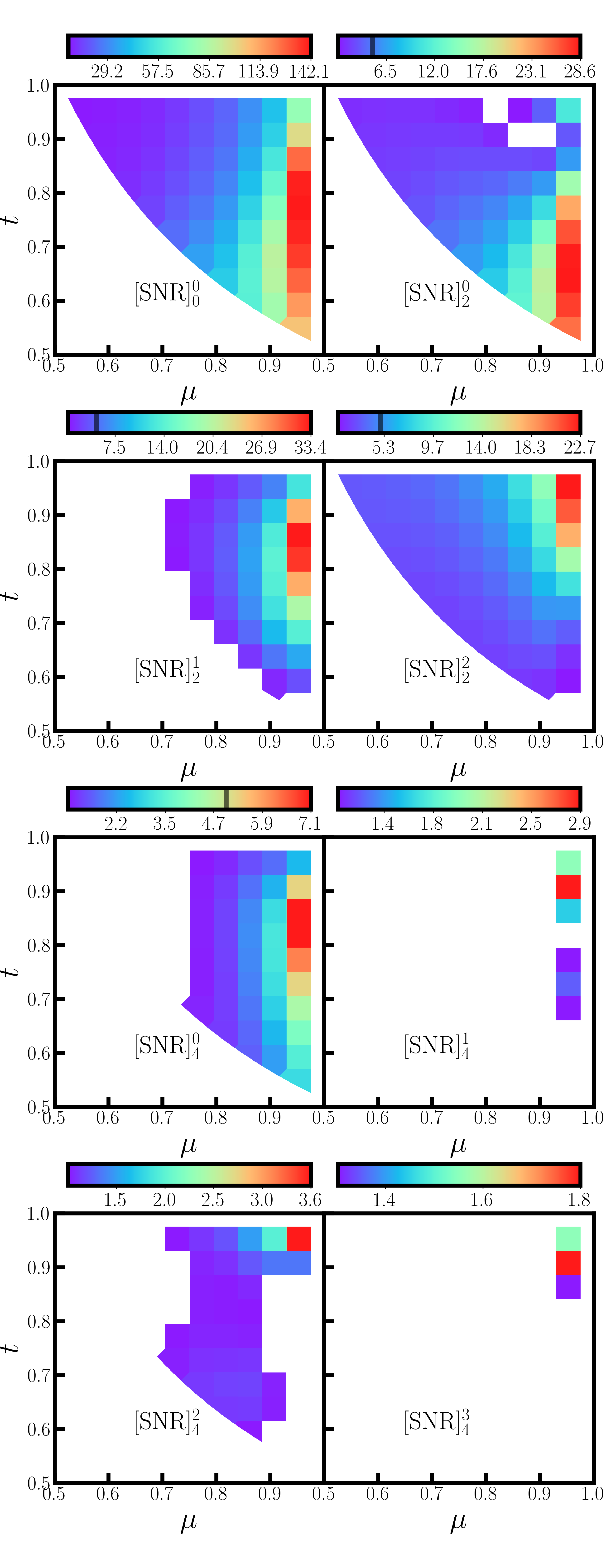}
    \caption{This shows the signal to noise ${[\rm SNR]}_\ell^m$ (eq.~\ref{eq:snr}) for different 
    multipoles with   $k_1 = 0.2 {\rm Mpc}^{-1}$ and $z=0.7$ fixed. 
    The regions where  ${[\rm SNR]}_\ell^m < 1$ have been masked out from the plots. 
    We do not show the multipoles for which ${[\rm SNR]}_\ell^m$ never exceeds unity.
    The vertical grey lines in the color bars mark the values ${[\rm SNR]}_\ell^m=5$.}
    \label{fig:snr_0.2}
\end{figure}

We now quantify the prospects of detecting the various bispectrum multipole moments $B_\ell^m$ using the 
signal-to-noise ratio (SNR) which is defined as 
\begin{eqnarray}\label{eq:snr}
{[\rm SNR]}_\ell^m = {|B_\ell^m|\over \sqrt{C_{\ell\ell}^{mm}}}\, .
\end{eqnarray}
A value ${[\rm SNR]}_\ell^m \ge 5$ (or possibly ${[\rm SNR]}_\ell^m \ge 3$) would be considered as a statistically significant detection of the particular  multipole moment.  As mentioned earlier, we have considered   bins of width  $(\Delta \ln k_1 = 0.1 ,\Delta \mu = 0.05, \Delta t = 0.05)$.  We note that the signal to noise ratio will increase if we consider bins  of larger widths.  Considering a situation where our analysis predicts ${[\rm SNR]}_\ell^m \approx 1 $,  it may still be possible to achieve a statistically significant detection by widening the bin width using $(\Delta \ln k_1 = 0.2 ,\Delta \mu = 0.1, \Delta t = 0.1)$ which still retains a significant amount of the information regarding the length-scale and shape dependence of the bispectrum. 
Further  the values of the various parameters used in our predictions are rather uncertain, and it is possible that the actual parameter values could result in  a larger  ${[\rm SNR]}_\ell^m$. Motivated by these factors,  we have shown the results for  ${[\rm SNR]}_\ell^m \ge 1$
where the signal and noise have equal amplitude, and we have also included this in the discussion.

\Fig{fig:snr_0.2} shows  ${[\rm SNR]}_\ell^m$ for different multipoles with   $k_1 = 0.2 {\rm Mpc}^{-1}$ and $z=0.7$ fixed. The regions where  ${[\rm SNR]}_\ell^m < 1$ have been masked out from the plots, and the multipoles where ${[\rm SNR]}_\ell^m$ never exceeds unity have not been shown.  Broadly, the ${[\rm SNR]}_\ell^m$ values decrease with increasing $\ell$, and for a fixed $\ell$  they  decrease with increasing $m$. However, we find an exception that the maximum value of ${[\rm SNR]}_2^1$ exceeds that of ${[\rm SNR]}_2^0$. For all ${[\rm SNR]}_\ell^m$
 we find that the maximum value  occurs for linear triangles $(\mu=1)$, however the value of $t$ corresponding to the maxima varies depending on $\ell$ and $m$.  Further, in most cases   ${[\rm SNR]}_\ell^m$ decreases as  the shape of the triangle is changed from linear $(\mu =1)$ to other  obtuse triangles $(\mu >t)$ and then acute triangles  $(\mu< t)$  and finally  the  equilateral triangle $(\mu,t) =(0.5,1)$.  
 Considering  $B_0^0$, we see that this can be detected at a high level of precision $({[\rm SNR]}_0^0>7)$ for all triangle configurations, and 
${[\rm SNR]}_0^0$ is maximum $(\approx 140)$ around $(1,0.75)$ . Considering  $B_2^0$, we see that the SNR  exceeds unity for most triangle configurations,  however ${[\rm SNR]}_2^0 < 5$ over a considerable region where $t > \mu$. We also find   a small region with ${[\rm SNR]}_2^0 < 1$  near the squeezed limit where the values of $B_2^0$ have a  zero  crossing (\fig{fig:qlm}). The maximum value  ${[\rm SNR]}_2^0 =28.6$ occurs around $(1,0.6)$ which is near the stretched limit. Considering ${[\rm SNR]}_2^1$, we see that the maximum value $(\approx  33.4)$ occurs around $(1,0.85)$ and we have ${[\rm SNR]}_2^1 >5$ in a region surrounding this. As mentioned earlier, the maximum value of ${[\rm SNR]}_2^1$ exceeds that of all the other multipole moments barring ${[\rm SNR]}_0^0$, however the condition ${[\rm SNR]}_2^1>1$ is satisfied for a relatively small range of shapes compared to ${[\rm SNR]}_2^0$ and ${[\rm SNR]}_2^2$. The maximum value of ${[\rm SNR]}_2^2$ ($\approx 22.7 $) occurs for squeezed triangles, and we have ${[\rm SNR]}_2^2>5$ in a region around this. The condition ${[\rm SNR]}_2^2>1$ is satisfied over most of the $(\mu,t)$ space except for a small region near the stretched triangles. 
${[\rm SNR]}_4^0$ is very similar to  ${[\rm SNR]}_2^1$, except that the values are smaller and the maximum value now is $\approx 7.1$. The higher multipoles with $\ell=4$ and $m=1,2$ and $3$ do not have ${[\rm SNR]}_{\ell}^m >5$ anywhere, however there is a rather large region where ${[\rm SNR]}_4^2>1$. The maxima of ${[\rm SNR]}_4^2$ $(\approx 3.6)$ occurs at the squeezed limit, whereas for $(m=1,3)$ the maxima $(2.9,1.8)$ occur very close to the squeezed limit. The two latter multipoles satisfy ${[\rm SNR]}_{\ell}^m>1$ in only a small region around the maxima.

\begin{figure*}
\centering
\includegraphics[width=1\textwidth]{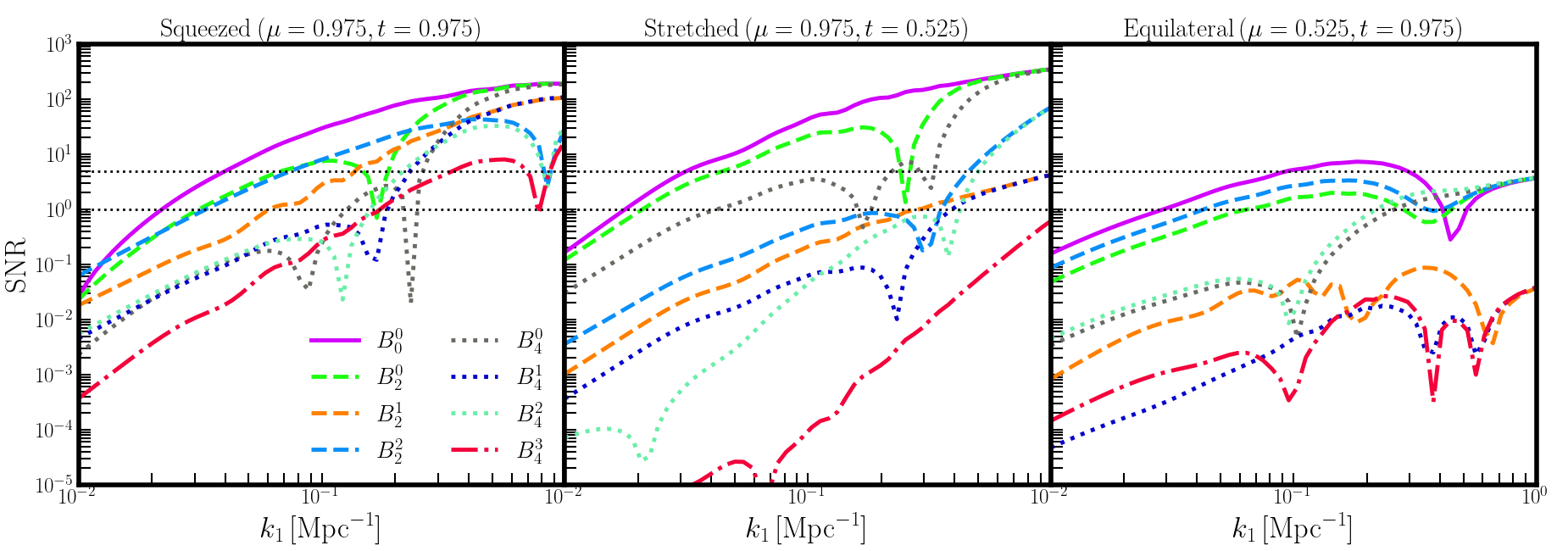}
\caption{This shows the variation of ${[\rm SNR]}_\ell^m$ (eq.~\ref{eq:snr}) 
with $k_1$, the largest wavevector (see fig.~\ref{fig:binning}), at fixed $z=0.7$.
We show results for three specific triangle shapes:
{\bf squeezed} (left-hand panel), {\bf stretched} (central panel) and 
{\bf equilateral} (right-hand panel).
The horizontal dotted-black lines show the ${[\rm SNR]}_\ell^m=1$ and $5$ levels
respectively. Note that we show ${[\rm SNR]}_\ell^m$ for the same multipoles as
in fig.~\ref{fig:snr_0.2}.}
\label{fig:snr_k}
\end{figure*}

We now consider the $k_1$ dependence of ${[\rm SNR]}_\ell^m$ with $z=0.7$ fixed,  as shown   in \fig{fig:snr_k}.  The smallest value of $k_1$ roughly corresponds to the linear extent of the survey.  The results are shown  for only three different triangle shapes  namely squeezed, stretched  and equilateral.   However, we can combine these with \fig{fig:snr_0.2} to qualitatively infer  the $k_1$ dependence expected for other triangle configurations. 
The number of triangles in any bin scales as  $N_{tr} \propto k_1^3$ (eq.~\ref{eq:ntr}), and we expect the cosmic variance to scale as $k_1^{-3}$   (eq.~\ref{eq:cov_def_3}) as $k_1$ is increased. Based on this we may expect  ${[\rm SNR]}_{\ell}^m$  to increase  monotonically with increasing $k_1$. This is broadly true for all the triangles in \fig{fig:snr_k}, except for the sharp dips which are seen to occur for  a few of the multipoles at certain values of $k_1$. As   seen in \fig{fig:qlm},  some of the multipoles have both positive and negative values, and the dips  in \fig{fig:snr_k} correspond to the zero crossings which are also reflected in  ${[\rm SNR]}_\ell^m$.  We now discuss the prospects of detecting the various multipoles at different values of $k_1$. Here we use $k_1[5]$ to denote the smallest value of $k_1$ where the condition ${[\rm SNR]}_{\ell}^m>5$  is satisfied, and generally we expect the $B_{\ell}^m$ to be detectable for all $k_1 \ge k_1[5]$.  Considering $B_0^0$ we see  that  $k_1[5] =  0.05  {\rm Mpc}^{-1}$ for the squeezed and stretched triangles, and  $k_1[5] =  0.1  {\rm Mpc}^{-1}$  for equilateral triangles.  Overall, we expect to detect $B_0^0$  for all  linear triangles and some  obtuse triangles at $k_1 \ge 0.05  {\rm Mpc}^{-1}$,   and  for all triangle  at $k_1 \ge 0.1  {\rm Mpc}^{-1}$.    Note that $B^0_0$  is predicted to be negative  at large $k_1$ for equilateral triangles. We find   a small $k_1$  range   around $k_1 \approx   0.4  {\rm Mpc}^{-1}$ where we do not expect to detect $B_0^0$ for equilateral and possibly some of the acute triangles.  Considering $B_2^0$, we have $k_1[5] = 0.05  {\rm Mpc}^{-1}$  and $ 0.1  {\rm Mpc}^{-1}$  for  stretched and squeezed triangles respectively.  The SNR falls off towards equilateral triangles where we have ${\rm SNR}  \ge 1 $ for $k_1 \ge  0.1  {\rm Mpc}^{-1}$ and $k_1[5] = 0.6  {\rm Mpc}^{-1}$.   Overall  we expect  $B_2^0$ to be detectable  in the vicinity of the stretched limit  for  $k_1 \ge  0.05  {\rm Mpc}^{-1}$, and for several other linear and obtuse triangles for   $k_1 \ge  0.1  {\rm Mpc}^{-1}$. Considering acute and equilateral triangles, it will be possible to detect  $B_2^0$  at $k_1 \ge  0.1  {\rm Mpc}^{-1}$ if we  increase the bin widths.  For all shapes,   $B_2^0$ is predicted to be negative at large $k_1$. 
The transition from positive to negative values occurs at  $k_1 \approx   0.1  {\rm Mpc}^{-1}$ for squeezed triangles, and then spreads to other shapes (stretched and then equilateral) as  $k_1$ is increased. As seen in  \fig{fig:snr_0.2}, it will not be possible to detect  $B_2^0$  in a few $\mu-t$ bins where the zero crossing  occurs. The exact $\mu-t$ locations of these bins   will shift with $k_1$. 
Considering $B_2^1$ and $B_2^2$ together, we see that   $k_1[5] \approx   0.1  {\rm Mpc}^{-1}$ for squeezed triangles, and we   expect to detect  these multipoles  for  several linear and obtuse  triangles  for 
 $k_1 \ge   0.1  {\rm Mpc}^{-1}$ . Further,  in all three cases (\fig{fig:snr_k})  $B^2_2$ satisfies ${\rm SNR} \ge 1$   for   $k_1 >   0.1  {\rm Mpc}^{-1}$.   In this $k_1$ range we expect   to detect  $B^2_2$  across   the entire $\mu-t$ plane if we consider larger bin widths.   We find that the rest of $B_{\ell}^m$  shown here ($\ell=4, m=0,1,2,3$) all  have $k_1[5]$ in the range $ 0.2-0.3   {\rm Mpc}^{-1} $
 for squeezed triangles, and  we   expect to detect  these multipoles  for  several linear and obtuse  triangles near the squeezed limit. We note that $B_4^0$ also has $k_1[5]$ in the range $ 0.2-0.3   {\rm Mpc}^{-1} $
 for stretched  triangles, and the region of $\mu-t$ space where this will be detected is relative large compared to the other $\ell=4$ multipoles.  We also note that $B^0_4$ and $B_4^2$ both satisfy ${\rm SNR} \ge 1$  for  the three triangle shapes shown in   \fig{fig:snr_k} for   $k_1 \ge   0.3  {\rm Mpc}^{-1}$, and we expect to detect these over the entire $\mu-t$ space if we increase the bin widths.

\begin{figure*}
\centering
\includegraphics[width=1\textwidth]{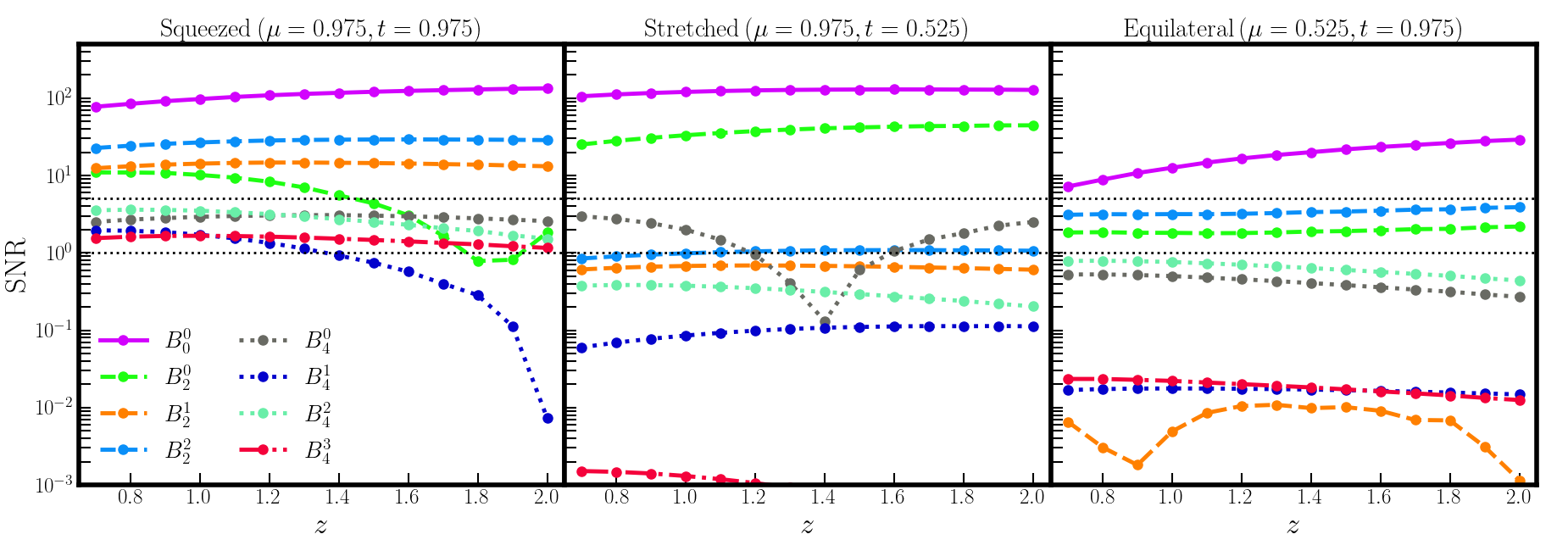}
\caption{This shows the variation of ${[\rm SNR]}_\ell^m$ (eq.~\ref{eq:snr}) 
with redshift at fixed $k_1=0.2\mpci$.
We show results for three specific triangle shapes:
{\bf squeezed} (left-hand panel), {\bf stretched} (central panel) and 
{\bf equilateral} (right-hand panel).
The horizontal dotted-black lines show the ${[\rm SNR]}_\ell^m=1$ and $5$ levels
respectively. Note that we show ${[\rm SNR]}_\ell^m$ for the same multipoles as
in fig.~\ref{fig:snr_0.2}. Filled circles represent the $z$ bins adopted 
from \citet{Yankelevich:2018uaz} to 
show our results.}
\label{fig:snr_z}
\end{figure*}

\Fig{fig:snr_z} shows how ${[\rm SNR]}_\ell^m$ varies with redshift with $k_1=0.2\mpci$ fixed.
Note that the bias parameters $b_1,\gamma_2$, as well as the survey volume $V$ change with changing
redshift, and we have used  the redshift dependent values from Table~1 of \citep{Yankelevich:2018uaz}.
We see that the survey volume $V$ increases with $z$,  and we expect the cosmic variance at a fixed $k_1$ to decrease with increasing $z$. As a result of this, ${[\rm SNR]}_\ell^m$ is expected 
to increase monotonically with redshift. However, there are two more effects that determine 
${[\rm SNR]}_\ell^m$ values at each $z$:
$(i)$ the amplitude of the bispectrum decreases with increasing $z$, and 
$(ii)$ the FoG suppression also decreases with increasing $z$. 
The bias parameters also affect the $z$ dependence of ${[\rm SNR]}_\ell^m$. 
We  see that  for many  of the multipoles ${[\rm SNR]}_\ell^m$  increase slightly with increasing $z$  whereas  there are some where the opposite occurs,  and for a few  cases we also have  dips in ${[\rm SNR]}_\ell^m$ at some values of $z$. As discussed earlier, these dips  correspond to the   zero crossings of $B_{\ell}^m$. 
We now discuss the prospects of detecting the different multipoles at the various redshifts shown here.  Considering $B_0^0$, we see that this is detectable with ${[\rm SNR]}_0^0>5$
at all the redshifts for all the three triangles considered. $B_2^0$ is detectable with ${[\rm SNR]}_2^0>5$ at all redshifts for stretched triangles. For squeezed triangles, the
above is true for $z<1.5$, whereas for equilateral triangles $1< {[\rm SNR]}_2^0<5$ at all redshifts. 
Considering $B_2^1$ and $B_2^2$, we see that they will be detected with ${[\rm SNR]}_\ell^m>5$ 
at all the redshifts only for squeezed triangles. For the stretched and equilateral triangles, 
${[\rm SNR]}_2^1$ and ${[\rm SNR]}_2^2$ values are below $5$ at all redshifts. 
All the $\ell>2$ multipoles cannot be observed with ${[\rm SNR]}_\ell^m>5$ at any redshift and
for the triangle shapes considered. Note that, the above discussion is true only for 
for three specific triangle shapes at  $k_1=0.2\mpci$. However, as we have discussed earlier, it is possible to detect the higher order multipoles with ${[\rm SNR]}_\ell^m>5$ by considering triangles of  other  shapes or  somewhat  larger $k_1$ values or larger bin widths. 

\subsection{Correlation between different multipoles}
\begin{figure}
    \centering
    \includegraphics[width=1\columnwidth]{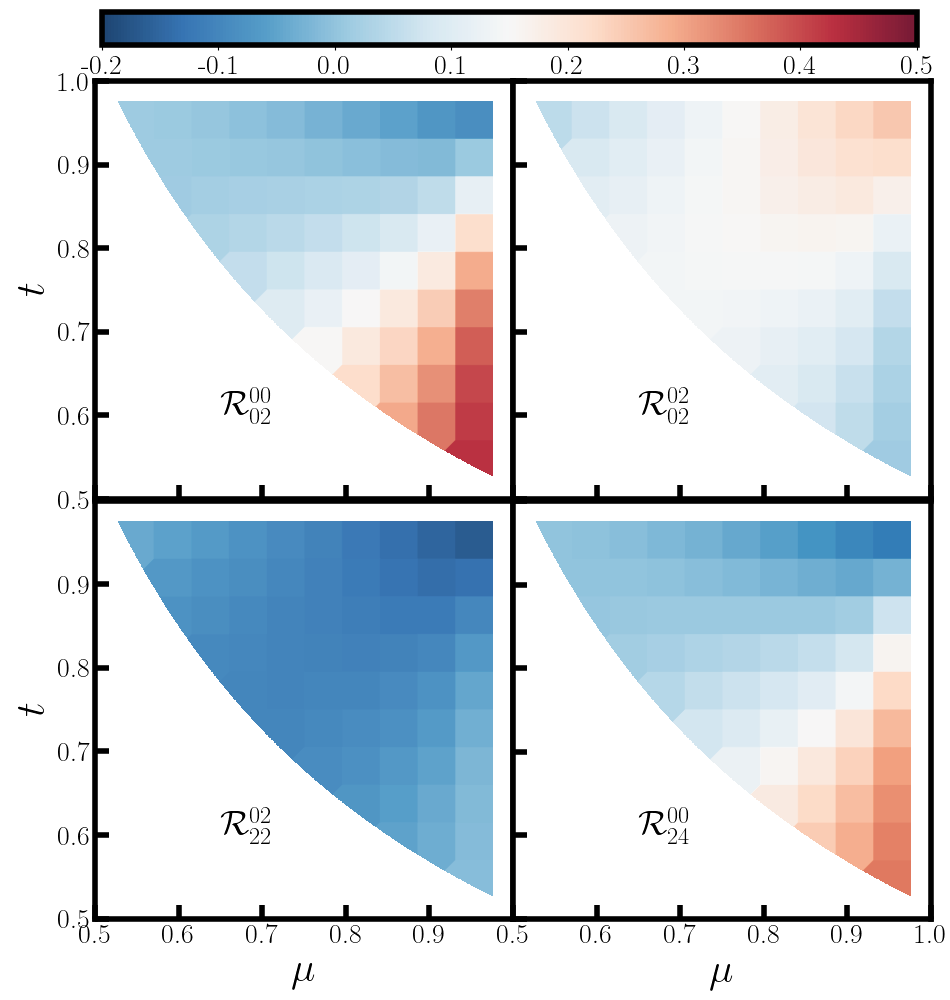}
    \caption{This shows rank correlations ${\cal R}_{\ell\ell'}^{mm'}$ 
    (eq.~\ref{eq:rank}) between different multipoles
    at fixed $k_1=0.2 {\rm Mpc}^{-1}$ and $z=0.7$. Note that, we do not show 
    rank correlations for which $\mid {\cal R}_{\ell\ell'}^{mm'} \mid <0.1$ 
    throughout the $\mu-t$ plane.}
    \label{fig:rank_0.2}
\end{figure}

It is important to note that the estimated values of the different multipole moments are expected to be correlated. We have quantified  the correlations in the measurement errors between any two multipoles using  the rank correlation
\begin{eqnarray}\label{eq:rank}
{\cal R}_{\ell\ell'}^{mm'} = { C_{\ell\ell'}^{mm'} \over \sqrt{C_{\ell\ell}^{mm}}\sqrt{C_{\ell'\ell'}^{m'm'}}} \,.
\end{eqnarray}
with $-1 \le  {\cal R}_{\ell\ell'}^{mm'} \le 1$.  Here positive and negative values indicate that the errors are correlated and anti-correlated respectively. 
We show the rank correlations between different multipoles in \fig{fig:rank_0.2}. 
We find that for most multipole pairs, the errors are only weakly correlated ($\mid {\cal R}_{\ell\ell'}^{mm'} \mid <0.1$ ) across much of the $\mu-t$ plane. However, there are a few exceptions which we highlight below. We first  consider   ${\cal R}_{02}^{00}$  which refers to $B_0^0$ and $B_2^0$ which have the highest  ${\rm SNR}$. We see that the  measurement errors in these two multipoles are correlated for obtuse triangles and anti-correlated for acute triangles.  ${\cal R}_{02}^{00}$ shows the highest correlation  $(0.42)$  in the stretched limit, and the highest anti-correlation  $(-0.1)$ in the squeezed limit. 
 Considering ${\cal R}_{02}^{02}$, we see that the  measurement errors of the  pair ($B_0^0$,$B_2^2$) are correlated with values  $\sim 0.25$  in the vicinity of the squeezed limit.  Considering ${\cal R}_{22}^{02}$, we see that this  show maximum anti-correlation $(-0.18)$ in the squeezed limit. ${\cal R}_{24}^{00}$ shows the highest correlation  $(0.33)$ and anti-correlation $(-0.1)$ in the stretched and squeezed limits respectively.  We have  $\mid {\cal R}_{\ell\ell'}^{mm'} \mid <0.1$ across the entire $\mu-t$ plane for all the other multipoles not shown here. Overall, the errors in the various multipoles are weakly correlated  for most triangle shapes barring a few in the vicinity of squeezed and stretched triangles.


\begin{figure*}
    \centering
    \includegraphics[width=1\textwidth]{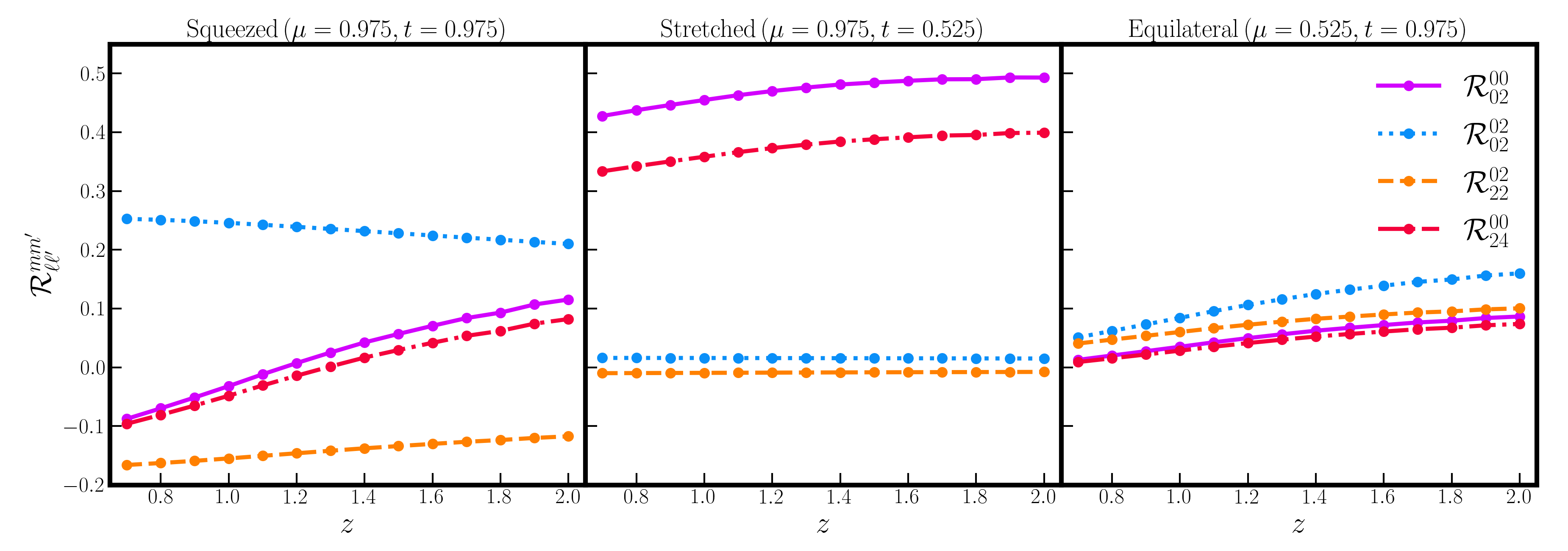}
        \caption{This shows the redshift dependence of the rank correlations 
        ${\cal R}_{\ell\ell'}^{mm'}$ (eq.~\ref{eq:rank}) between different multipoles
        for a fixed $k_1=0.2\mpci$. We show results for three specific triangle shapes:
        {\bf squeezed} (left-hand panel), {\bf stretched} (central panel) and 
        {\bf equilateral} (right-hand panel).
        Note that, we have considered the same rank correlations as shown in \fig{fig:rank_0.2}.
        Filled circles represent the $z$ bins adopted 
        from \citet{Yankelevich:2018uaz} to 
        show our results.
        }\label{fig:rank_z}
\end{figure*}

\Fig{fig:rank_z} shows the redshift dependence (from  $z=0.7$ to $2$) of the four rank correlations  shown in \fig{fig:rank_0.2} for three specific triangle shapes with  $k_1=0.2\mpci$ fixed.  Considering  ${\cal R}_{02}^{00}$ (${\cal R}_{24}^{00}$) we see that  the peak correlation, which occurs for  stretched triangles,  increases from $0.42$  ($0.32$)  to $\sim 0.5$ ($0.4$)    whereas the value changes from $\sim -0.1$ to $\sim 0.1$ for squeezed triangles.  Considering  ${\cal R}_{02}^{02}$, the peak correlation, which occurs for  squeezed  triangles,   drops from $\sim 0.25$ to $0.2$ across  $z$.  Considering  ${\cal R}_{22}^{02}$,  for squeezed triangles its value  changes from $-0.18$ to $-0.11$ across $z$.  The other cases shown in the figure all have $\mid {\cal R}_{\ell\ell'}^{mm'} \mid <0.1$, with the exception of  ${\cal R}_{02}^{02}$ which has values in the range  $0.1$ to $0.2$ at  $z \ge 1$ for equilateral triangles.

\begin{figure*}
    \centering
    \includegraphics[width=1\textwidth]{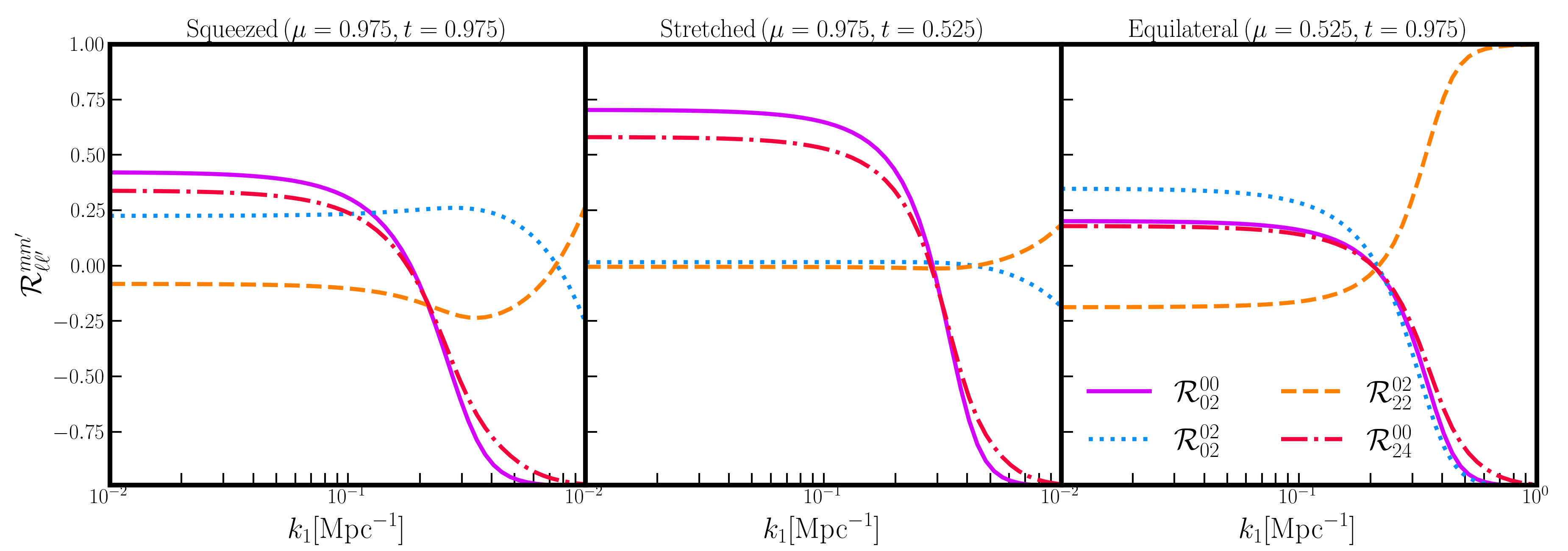}
    \caption{This shows the $k_1$ dependence of the rank correlations 
        ${\cal R}_{\ell\ell'}^{mm'}$ (eq.~\ref{eq:rank}) between different multipoles
        for a fixed $z=0.7$. We show results for three specific triangle shapes:
        {\bf squeezed} (left-hand panel), {\bf stretched} (central panel) and 
        {\bf equilateral} (right-hand panel).
        Note that, we have considered the same rank correlations as shown in \fig{fig:rank_0.2}.}
    \label{fig:rank_k}
\end{figure*}

\Fig{fig:rank_k} shows the $k_1$ dependence of the four rank correlations shown in  \fig{fig:rank_0.2} for for three specific triangle shapes with  $z=0.7$ fixed. Considering all the panels, we see that the rank correlations are practically independent of 
  $k_1$ for $k_1 \lesssim 0.05\mpci$. We have already mentioned that
 the FoG effect can be neglected at $k_1 \lesssim 0.05 \mpci$. The  numerator and denominator of eq.~(\ref{eq:rank}) are both dominated by the power spectrum contribution, and the 
 $k_1$ dependent  factors cancel out to  make ${\cal R}_{ll'}^{mm'}$  independent of $k_1$ 
 in the absence of the FoG effect. However, we  see a strong $k_1$ dependence for $k_1 > 0.1 \mpci$ where the FoG is important. We see that  ${\cal R}_{02}^{00}$ and ${\cal R}_{24}^{00}$  exhibit very similar $k_1$ dependence in all the three panels where we have positive correlations at $k_1<0.05 \mpci$, the correlations decline with increasing $k_1$ and cross  zero  around  $k_1\sim0.2$    and we have nearly complete anti-correlation  $(\sim -1)$ for  $k_1 >0.5 \mpci$. Considering  $k_1<0.05 \mpci$ we have ${\cal R}_{02}^{00} >{\cal R}_{24}^{00}$,  with  the  maximum correlation   $(0.5-0.75)$  occurring  for stretched triangles, followed by values in the range  $0.3-0.5$ for  squeezed  triangles and  $0.2-0.3$ for equilateral triangles.  The $k_1$ dependence of ${\cal R}_{02}^{02}$ is very similar to that of ${\cal R}_{02}^{00}$ and $ {\cal R}_{24}^{00}$ for equilateral triangles. For squeezed (stretched)  triangles  ${\cal R}_{02}^{02} \sim 0.25 \, (0.0)$ for $k_1  \le 0.4 \mpci $ beyond which both  sharply declines to $\sim -0.25$ at  $k_1 = 1 \mpci $. In contrast, ${\cal R}_{22}^{02}$ shows exactly the opposite $k_1$ dependence in that we have anti-correlations for small $k_1$ and correlations at large $k_1$, with a zero crossing whose $k_1$ value depends on the shape of the triangle.  Considering squeezed, stretched and equilateral triangles  we respectively have ${\cal R}_{22}^{02} \sim -0.1,0.0$ and $-0.2$ at small $k_1$  $(\le 0.1 \mpci)$ , whereas these values are $\sim 0.25,0.25$ and $1.0 $ for large $k_1$   $(\le 0.1 \mpci)$.

\section{Discussion and Conclusion}\label{se:conclusion}

The redshift space anisotropy of the bispectrum is generally quantified using multipole moments as defined in  
eq.~(\ref{eq:bispec_int}). The possibility of measuring these multipoles in any survey 
depends on the level of statistical fluctuations.
In this paper, we developed a formalism to compute the statistical 
fluctuations in the measurement of bispectrum multipoles for galaxy surveys. 
We quantify the fluctuations through the covariance as defined in eq.~(\ref{eq:cov_def_3}) which 
assumes the flat sky approximation. We consider the specifications of a {\it Euclid} like galaxy 
survey to present our results. We mainly consider two quantities: 
$(i)$ the signal-to-noise ratio(SNR) ${[\rm SNR]}_\ell^m$ which quantifies the detectability of a multipole
$B_\ell^m$ (eq.~\ref{eq:snr}), and
$(ii)$ the rank correlation which quantifies the 
correlation in measurement errors between any two multipoles (eq.~\ref{eq:rank}).
We show, how these quantities depend on the triangle configurations in ${\bf k}$ space,
as well as their evolution in redshift. We also show how our results change as we
introduce the FoG effect.

We find that the FoG effect plays a crucial role at length scales $k>0.05\mpci$ for $z=0.7$. 
This suppresses the values of bispectrum multipoles and the suppression is in general stronger 
for the higher order multipoles. The amplitude of the monopole $B_0^0$ predicted by 2PT
is reduced by half when FoG effect is introduced. For $B_4^m$, the amplitude
is suppressed by almost an order of magnitude in presence of FoG effect. 
This FoG suppression, however, is not very important at small $k$ or at high $z$.

We see that in general ${[\rm SNR]}_\ell^m$ values decrease with increasing $\ell$, 
and for a fixed $\ell$  they  decrease with increasing $m$. 
Considering all ${[\rm SNR]}_\ell^m$, we find that the maximum value occurs for 
linear triangles $(\mu=1)$, however the value of $t$ corresponding to the maxima varies 
depending on $\ell$ and $m$. Also, in most of the cases ${[\rm SNR]}_\ell^m$ decreases 
as the shape of the triangle is changed from linear $(\mu =1)$ to 
other obtuse triangles $(\mu >t)$ and then acute triangles  $(\mu< t)$  
and finally the  equilateral triangle $(\mu,t) =(0.5,1)$. We note that, at large,
${[\rm SNR]}_\ell^m$ increases with $k_1$ and $z$.

Considering individual multipoles, we expect to detect $B_0^0$ for all the triangles 
at $k_1>0.1 \mpci$ across the redshifts considered here. On the other hand, we expect 
$B_2^0$ to be detectable in the vicinity of the stretched limit for $k_1>0.05 \mpci$ at
all the redshifts, and for several other linear and obtuse triangles for $k_1 \geq 0.1$
up to $z\sim 1.5$. Detection of $B_2^0$ for acute and equilateral triangles, as well as 
for obtuse triangles at $z>1.5$, is possible if we increase the bin size. Detection of $B_2^1$ and 
$B_2^2$ at all redshifts is possible only for squeezed triangles at $k_1\geq 0.1 \mpci$.
For other triangle shapes, detection is possible at higher $k_1$. Considering $B_4^m$ multipoles,
we see that the detection is possible across the redshift range only for a few linear and obtuse 
triangles near the squeezed limit at $k_1>0.2 \mpci$. Note that, for all the multipoles, 
the possibility of detection, or the ${[\rm SNR]}_\ell^m$ values, can be increased
by increasing the bin size in $k_1, \mu$ or $t$.

Considering ${\cal R}_{\ell\ell'}^{mm'}$, we find that for most multipole pairs the errors are only weakly correlated (with $\mid {\cal R}_{\ell\ell'}^{mm'} \mid <0.1$ ) across much of the $\mu-t$ plane
barring a few in the vicinity of squeezed and stretched triangles. For a fixed triangle shape, 
the ${\cal R}_{\ell\ell'}^{mm'}$ values are practically independent of $k_1$ for $k_1 \lesssim 0.05 \mpci$
and show a strong $k_1$ dependence for $k_1>0.1 \mpci$ where FoG effect is really important. For a fixed
triangle shape and $k_1$, ${\cal R}_{\ell\ell'}^{mm'}$ evolve moderately with redshift.

We, therefore, conclude that the future surveys like \euclid{} can potentially measure the higher order
redshift space bispectrum multipoles (up to $\ell=4$), 
beyond the isotropic component (monopole), across 
various triangle shapes and sizes. The signal-to-noise or significance of these measurements, however,
depend on the scales and redshifts of observation. The signal-to-noise also determines 
the information content of the individual multipoles. 
Significant measurements of $\ell \leq 2$ multipoles are possible even at 
$k_1 \lesssim 0.1 \mpci$ across the $z$ range $0.7 - 2$.
These scales are particularly important as the FoG suppression is minimum here.
Here, we expect to measure the $\ell \leq 2$ multipoles with highest signal-to-noise
for linear and obtuse triangles. For $\ell>2$ multipoles, we require to go to large 
$k_1$ for significant detection or we need large bin size to increase the 
signal-to-noise. Due to the weak correlation (at $k_1<0.2 \mpci$) between the errors of multipole pairs
for most of the triangle shapes, it is possible to combine different multipoles to
increase the information content. This becomes particularly important when we try to extract
cosmological parameters from bispectrum measurements. 
Following our analysis, we expect to reduce the errors on the
cosmological parameters when we combine the higher multipoles along with the 
monopole results \citep{Gil-Marin:2014sta,Gil-Marin:2016wya,Philcox:2021kcw}. 
This we plan to study in future. 
Finally, we reiterate that in \euclid{} like surveys, we expect to measure 
bispectrum multipoles up to $\ell = 4$ by 
suitably choosing the scale and redshifts of observation.

\section*{DATA AVAILABILITY}
The data generated during this work will be made available
upon reasonable request to the authors.

\bibliographystyle{mnras}

\input{main.bbl}

\end{document}